\documentclass[%
superscriptaddress,
reprint,
amsmath,amssymb,
aps,
prl,
longbibliography,
noeprint
]{revtex4-2}

\usepackage[pdftex]{graphicx}% Include figure files
\graphicspath{{C:/Users/scott/OneDrive/document/kosakaLab/paper/NV0_control_milliKelvin/figures/}}
\usepackage{braket}
\usepackage{here}
\usepackage{siunitx}
\usepackage{comment}
\usepackage{dcolumn}% Align table columns on decimal point
\usepackage{bm}% bold math
\usepackage{hyperref}% add hypertext capabilities
\usepackage{amsmath}
\usepackage{color}

\begin{document}
	
	\title{Quantum Orbital-State Control of a Neutral Nitrogen-Vacancy Center at Millikelvin Temperatures}

	\author{Hodaka Kurokawa}
	\email[E-mail: ]{kurokawa-hodaka-hm@ynu.ac.jp}
	\address{Quantum Information Research Center, Institute of Advanced Sciences, Yokohama National University, 79-5 Tokiwadai, Hodogaya, Yokohama, 240-8501, Japan}
	\author{Shintaro Nakazato}
	\address{Department of Physics, Graduate School of Engineering Science, Yokohama National University,  79-5 Tokiwadai, Hodogaya, Yokohama, 240-8501, Japan }
	\author{Toshiharu Makino}
	\address{Quantum Information Research Center, Institute of Advanced Sciences, Yokohama National University, 79-5 Tokiwadai, Hodogaya, Yokohama, 240-8501, Japan}
	\address{Advanced Power Electronics Research Center, National Institute of Advanced Industrial Science and Technology, 1-1-1 Umezono, Tsukuba, Ibaraki, 305-8568, Japan }
	\author{Hiromitsu Kato}
	\address{Quantum Information Research Center, Institute of Advanced Sciences, Yokohama National University, 79-5 Tokiwadai, Hodogaya, Yokohama, 240-8501, Japan}
	\address{Advanced Power Electronics Research Center, National Institute of Advanced Industrial Science and Technology, 1-1-1 Umezono, Tsukuba, Ibaraki, 305-8568, Japan }
	\author{Shinobu Onoda}
	\address{Quantum Information Research Center, Institute of Advanced Sciences, Yokohama National University, 79-5 Tokiwadai, Hodogaya, Yokohama, 240-8501, Japan}
	\address{Quantum Materials and Applications Research Center (QUARC),National Institutes for Quantum Science and Technology (QST), 1233, Watanuki-machi, Takasaki, Gunma, 370-1292, Japan}
	\author{Yuhei Sekiguchi}
	\address{Quantum Information Research Center, Institute of Advanced Sciences, Yokohama National University, 79-5 Tokiwadai, Hodogaya, Yokohama, 240-8501, Japan}
	\author{Hideo Kosaka}
	\email[E-mail: ]{kosaka-hideo-yp@ynu.ac.jp}
	\address{Quantum Information Research Center, Institute of Advanced Sciences, Yokohama National University, 79-5 Tokiwadai, Hodogaya, Yokohama, 240-8501, Japan}
	\address{Department of Physics, Graduate School of Engineering Science, Yokohama National University,  79-5 Tokiwadai, Hodogaya, Yokohama, 240-8501, Japan }
		
\begin{abstract}
A neutral nitrogen-vacancy center (NV$^0$) is promising for realizing strong coupling with a single microwave photon due to its large electric field sensitivity, although it is  susceptible to environmental phonon noise at 5 K. Decreasing the temperature to 15 mK results in a tenfold increase in orbital relaxation time compared to that at 5 K. Dynamical decoupling pulses significantly increase the orbital coherence time to around 1.8 $\si[per-mode=symbol]{\micro \second}$, representing a 30-fold improvement compared to that without decoupling pulses. Based on these results, a single NV$^0$ can reach the strong coupling regime when coupled with a high-impedance microwave resonator, thus  opening up  the possibility of microwave quantum electrodynamics using a single optically-active defect center in diamond.

\end{abstract}
\maketitle
	
%\vspace{10mm}
%\section{Introduction}
%introduction

%explanation about NV0 
%(experimental method)
%(energy level, PLE, ODER)

%Temperature dependence of T1

%Rabi, Ramsey
%Echo, CPMG
%discussion

Diamond color centers have been intensively studied due to their potential applications as quantum sensors \cite{Schirhagl2014},  single-photon sources \cite{Beveratos2002, Babinec2010,Mizuochi2012,Knall2022}, quantum processors \cite{Nemoto2014,Pezzagna2021,Abobeih2022,Sekiguchi2022}, quantum repeaters \cite{Sekiguchi2017,Pompili2021,Bhaskar2020,Knaut2024,Stolk2024}. Among these color centers, the negatively charged nitrogen-vacancy center (NV$^-$) is the most popular and well-studied due to its high quantum efficiency and the ability to operate at room temperature  \cite{Manson2005,Jelezko2006,Childress2006,Dolde2011,Pezzagna2021,Ruf2021}. Recently, however, group IV color centers characterized by higher zero-phonon line emission efficiency and better spectral stability in a nanostructure have gained increasing attention in applications for quantum communication \cite{Mu2014,Rogers2014a,Becker2018a,Iwasaki2017,Haußler2017,Bradac2019,Chen2020b, Ruf2021,Parker2024,Senkalla2024}. Research into novel diamond color centers for specific applications are still actively underway.

A neutral nitrogen-vacancy center, NV$^0$, has  an orbital degree of freedom in its ground state, resulting in a five orders of magnitude larger electric field \cite{Kurokawa2024} and strain \cite{Venturi2019} susceptibility compared to the ground state of NV$^-$ \cite{Oort1990,Udvarhelyi2018}. Since the large susceptibility to environmental phonons results in short orbital relaxation time (several hundred nanoseconds at 5 K) \cite{Baier2020}, NV$^0$ has not attracted as much interest as NV$^-$. However, its large electric field response and ground state splitting around 10 GHz \cite{Baier2020, Kurokawa2024} are advantageous when considering the orbital state control using microwave electric fields and realization of microwave cavity quantum electrodynamics using a single color center in diamond. While strong coupling to an optical photon has been achieved in a silicon vacancy center \cite{Nguyen2019}, and strong coupling to flux qubit has been realized with an ensemble of NV$^-$ centers \cite{Zhu2011}, strong coupling to a microwave field using a single color center remains a challenge.

In this study, we perform quantum control of the orbital state of NV$^0$ and estimate the orbital relaxation and coherence times at millikelvin temperatures. We observe an exponential increase in the orbital relaxation time due to a decrease in environmental phonons. Additionally, the orbital coherence is extended through the application of dynamical decoupling techniques. Based on the obtained relaxation and coherence times, strong coupling between NV$^0$ and a high-impedance microwave resonator \cite{Samkharadze2016,Niepce2019a} is achievable, suggesting the possibility of quantum electrodynamics experiments using an optically-active diamond color center, such as controlling and reading out NV$^0$ using a single microwave photon as well as optical photons. %Since it is in principle possible to incorporate NV$^0$ into an optical nanocavity to achieve strong coupling with an optical photon, this result paves the way for using NV$^0$ as a quantum interface between microwave and optical photons.

%Experiments
In the following, all measurements are performed in a dilution refrigerator using a homebuilt confocal microscopy system. The temperature at the mixing chamber plate, $T_\mathrm{MXC}$, is maintained at 50 mK unless specified. A diamond sample is mounted onto a stack of 3-axis piezo positioners and a 3-axis piezo scanner. A 10-$\si[per-mode=symbol]{\micro \meter}$-diameter solid-immersion lens is fabricated on the surface of the diamond to enhance the collection efficiency. Driving electric fields are applied using Au/Ti electrodes formed on the sample (See Supplementary Methods).

%%%
%Energy level structure of NV0

\begin{figure}
	\centering
		\includegraphics[width=85 mm]{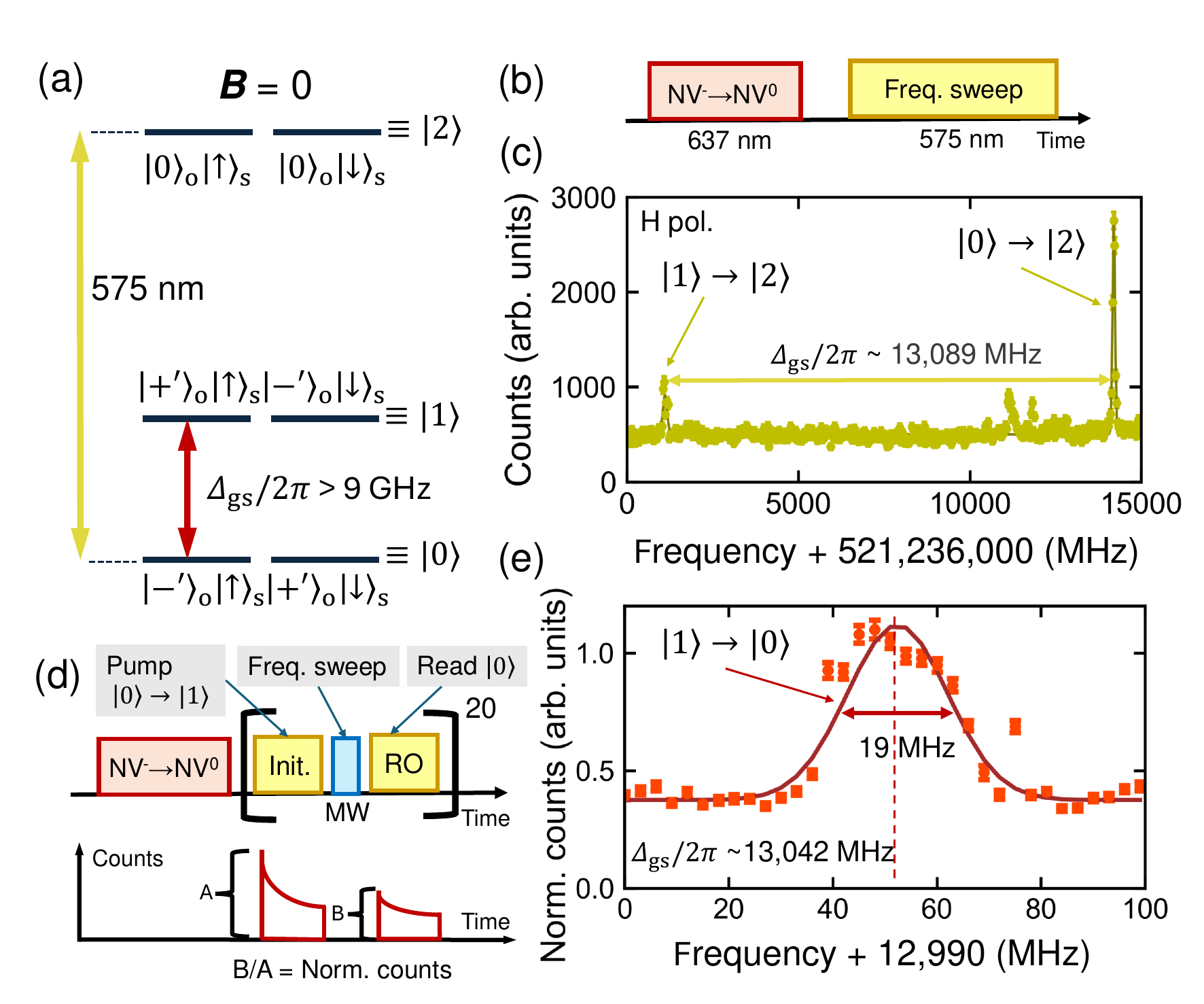}
	\caption{Energy level structure of NV$^0$ and its characterization. (a) Schematic of the energy levels of NV$^0$ with some static strain under a zero magnetic field. The ground state exhibits energy splitting on the order of 10 GHz. The optical transition occurs at 575 nm. $\ket{\pm}_\textrm{o}$ and $\ket{\uparrow}_\textrm{s}$ are the eigenstate of the orbital (o) and spin (s) operators. The prime symbol ($\ket{\pm'}$) indicates the orbital eigenstates are hybridized under the static strain \cite{Kurokawa2024}. 		 $\ket{0}_\textrm{o}$ represents that the state does not have the orbital angular momentum. (b) Measurement sequence for PLE spectrum of NV$^0$. The 637-nm laser that is resonant with the zero-phonon line of NV$^-$ is used to initialize the charge state to NV$^0$, while the 575-nm laser is used to observe the optical transition. (c) PLE spectrum of NV$^0$. The 575-nm laser is horizontally polarized to maximize counts from $\ket{0}$. (d) Measurement sequence for ODER. After the charge initialization, the driving microwave pulse is applied between initialization (Init.) and readout (RO) optical pulses. The initialization pulse pumps the orbital state from $\ket{0}$ to $\ket{1}$. Changes in population induced by the driving pulse are estimated from the ratio of the maximum counts (B/A) obtained from the first (A) and second (B) pulses. In the following analysis, the dark counts are subtracted from the PL counts. The PL counts are then smoothed by the five-point moving average. (e) The ODER spectrum corresponding to the  $\ket{1}\rightarrow\ket{0}$ transition, which is fitted by a Gaussian curve. The error bars correspond to the standard deviation considering the photon shot noise. 
	}
	\label{fig:energy-level}
\end{figure}

First, we briefly introduce the energy level structure of NV$^0$ and show experiments characterizing the energy level of NV$^0$. In the ground state, NV$^0$ has four energy levels with orbital ($\ket{\pm}_\textrm{o}$) and spin ($\ket{\uparrow}_\textrm{s}$, $\ket{\downarrow}_\textrm{s}$) degrees of freedom (Fig.\ref{fig:energy-level}(a)), which is the eigenstate of the Hamiltonian, $2\lambda \hat{L_\mathrm{z}}\hat{S_\mathrm{z}}$, where $\ket{\pm}_\textrm{o}=\mp(\ket{e_x}_\mathrm{o}\pm i\ket{e_y}_\mathrm{o})$, $\ket{e_x}_\mathrm{o}$ and $\ket{e_y}_\mathrm{o}$ are the orbital states of the nitrogen vacancy \cite{Maze2011,Baier2020}, $\hat{L_\mathrm{z}}=\hat{\sigma}_\textrm{z}$ is the orbital operator for $\ket{\pm}_\textrm{o}$ basis and $\hat{S_\mathrm{z}}=\hat{\sigma}_\textrm{z}/2$ is the half-spin operator \cite{Kurokawa2024}. Considering the static strain term in the Hamiltonian, $\epsilon_{\perp}(\hat{L}_+ + \hat{L}_-)$, the spin-orbit coupling, $\lambda$, and the strain, $\epsilon_{\perp}$, cause the ground-state splitting to the spin-orbital state as, $\Delta_\mathrm{gs}=2\sqrt{\lambda^2+\epsilon_{\perp}^2}$ \cite{Barson2019,Baier2020,Kurokawa2024}, in its ground state. Here, $\hat{L}_{\pm}$ is the orbital raising and lowering operators, which hybridize $\ket{\pm}_\mathrm{o}$.
 Under an ambient magnetic field, the spin states are degenerate. In the following experiments, we ignore the spin degree of freedom and treat it as an effective three-level system. We denote the lower energy level as $\ket{0}$ and the upper energy level as $\ket{1}$, and the optically excited level as $\ket{2}$.

Regarding the optical properties of NV$^0$, the optical excitation occurs under illumination with a 575-nm laser. After irradiating of a 637-nm laser resonant with the zero-phonon line of NV$^-$ for charge initialization, the frequency of the 575-nm laser is swept to obtain a photoluminescence excitation (PLE) spectrum (Fig.\ref{fig:energy-level}(b)).  The two observed peaks are identified as $\ket{0}\rightarrow\ket{2}$ and $\ket{1}\rightarrow\ket{2}$ transitions, respectively. The separation between two peaks corresponds to $\Delta_\mathrm{gs}$, which is 13,089 MHz (Fig. \ref{fig:energy-level}(c)). Assuming that $\lambda/2\pi=$4.9 GHz \cite{Baier2020}, $\epsilon_{\perp}/2\pi=4.3$  GHz for the measured NV$^0$. The minimum PLE linewidth (FWHM) is 44.8 MHz (See Supplementary Notes for detail), which is approximately six times larger than the Fourier transform-limited linewidth (of 7.2 MHz \cite{Baier2020}).

We also perform an optically-detected electrical resonance (ODER) experiment to confirm the ground-state resonance frequency using driving microwave electric fields as shown in Fig.\ref{fig:energy-level} (d).  After the charge initialization, the spin-orbital state is prepared by the optical pumping from $\ket{0}$ to $\ket{1}$. A 30-ns microwave pulse is applied to invert the population around the resonance frequency. Then, the population at $\ket{0}$ is read out by applying the second pulse. Both the first and the second pulses are resonant to the  $\ket{0}\rightarrow\ket{2}$ transition. The  photoluminescence (PL) during the measurements is recorded using a time-resolved photon counter with a 1.28 ns time bin. %The maximum height of the PL counts during the first and second pulses are assumed to be proportional to the population in $\ket{0}$ (Fig. \ref{fig:energy-level}(d)). 
The maximum PL counts of the second pulse are normalized to the maximum counts of the first pulse, which corresponds to changes in population during the first and second pulses.  %

The resulting ODER spectrum is shown in Fig. \ref{fig:energy-level}(e). The center is 13,042 MHz (=$\Delta_\mathrm{gs}/2\pi$) and the FWHM is 19$\pm0.4$ MHz from the Gaussian fit. $\Delta_\mathrm{gs}$ is in good agreement with the value obtained from the PLE measurement. The slight deviation in the estimation of $\Delta_\mathrm{gs}$ between PLE and ODER is attributed to the limited frequency resolution in the PLE measurement and the frequency drift during the measurement. The following measurement reveals that the Rabi frequency for this measurement is 14 MHz, limiting the ODER linewidth by the power broadening. However, due to spectral hopping, it is difficult to further reduce the ODER linewidth by decreasing the driving microwave power.

\begin{figure}
	\centering
		\includegraphics[width=75 mm]{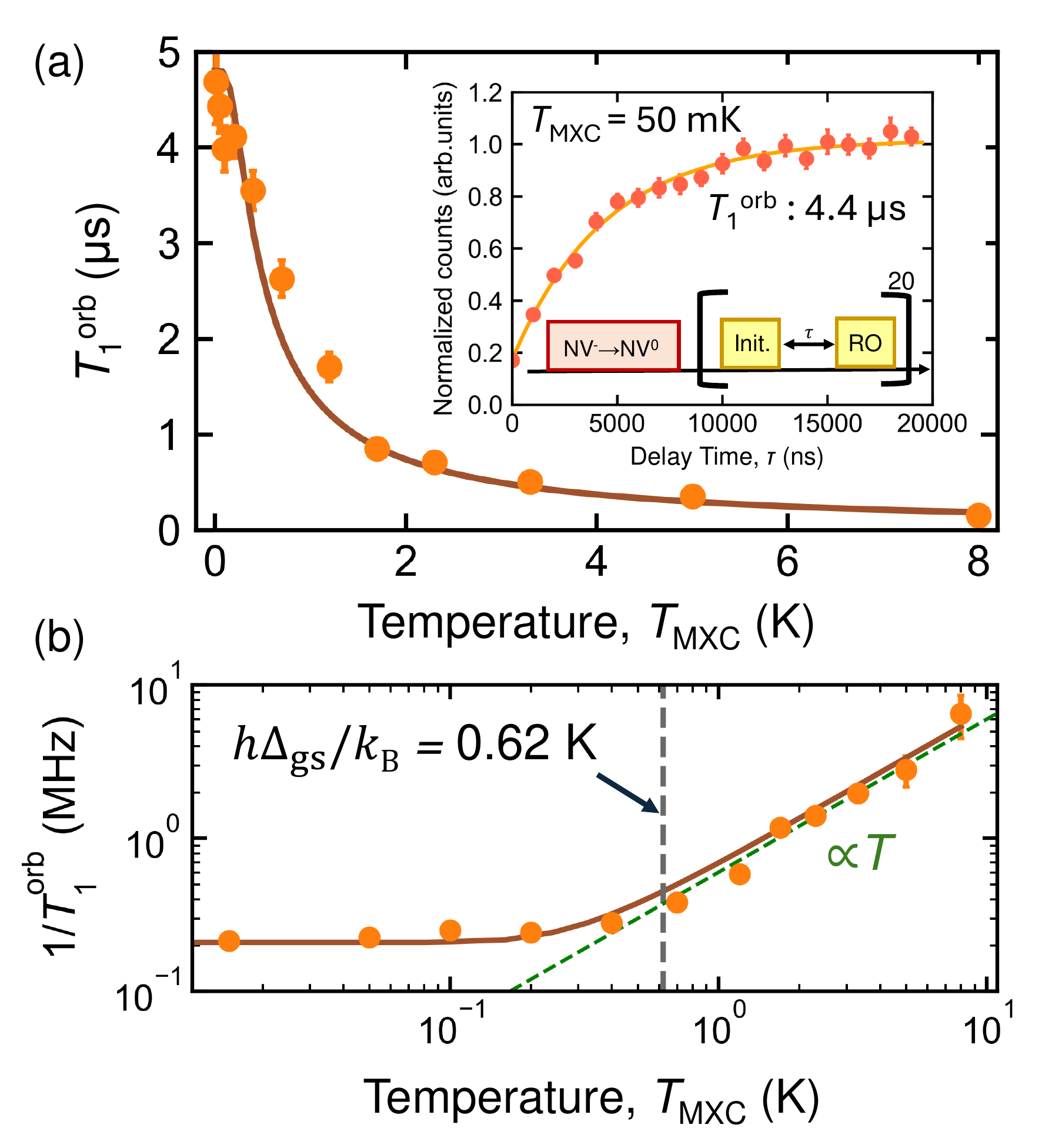}
	\caption{Temperature dependence of the orbital relaxation time and rate. (a) Temperature dependence of the orbital relaxation time. The fitting function for $T_1^{\mathrm{orb}}$ is shown in the main text. The inset shows how $T_1^{\mathrm{orb}}$ is extracted from the measurement of the population recovery after the pumping pulse from $\ket{0}$ to $\ket{1}$ at $T_\textrm{MXC}=$ 50 mK, where $T_\textrm{MXC}$ is the temperature of the mixing chamber plate. The error bars correspond to the standard error in the nonlinear fitting. The measurement sequence is also shown in the inset, representing that the time delay between the initialization and readout pulses is swept. (b) Temperature dependence of the orbital relaxation rate. The green dashed line is proportional to temperature. 
	}
	\label{fig:T1-temp}
\end{figure}

At millikelvin temperatures, the orbital relaxation time, $T_1^{\mathrm{orb}}$, of NV$^0$ is expected to increase compared to that at 5 K since phononic and electric excitations around 10 GHz in the environment are reduced by  several orders of magnitude. $T_1^{\mathrm{orb}}$ is estimated using the pump-probe technique similar to the ODER measurement. As expected, $T_1^{\mathrm{orb}}$ increases from 150$\pm$60 ns at $T_{\mathrm{MXC}}$=8.0 K to 4.7$\pm$0.4 $\si[per-mode=symbol]{\micro \second}$ at $T_{\mathrm{MXC}}$=10-15 mK (Fig.\ref{fig:T1-temp}(a)). The exponential increase in $T_1^{\mathrm{orb}}$ is in reasonably good agreement with a theoretical curve assuming a relaxation via a single-phonon process, $1/T_1^{\mathrm{orb}}=A\Delta_{\mathrm{gs}}^3(2n_{\mathrm{BE}}(\Delta_{\mathrm{gs}},T)+1)$ \cite{Jahnke2015}, where $A$ is a constant and $n_{\mathrm{BE}}$ is the average population in the Bose-Einstein distribution. The reasonable agreement with the fitting curve indicates that $T_\mathrm{MXC}$ is close to the sample temperature. The exponential increase in $T_1^{\mathrm{orb}}$ is explained by the existence of the ground-state splitting (energy gap). When the ground-state splitting exceeds the thermal energy, $k_\textrm{B}T$, with decreasing temperature, thermal excitation is suppressed exponentially, resulting in the increase in $T_1^{\mathrm{orb}}$. At $T_\mathrm{MXC} >$ 0.62 K, $1/T_1^{\mathrm{orb}}$ increases linearly with  temperature as shown in Fig. \ref{fig:T1-temp}(b), indicating that the single-phonon process is dominant. The linear temperature dependence at higher temperatures is consistent with the report in Ref. \cite{Baier2020}. It should be noted that  $T_1^{\mathrm{orb}}$ is several orders of magnitude smaller than the spin relaxation time, $T_1^{\mathrm{spin}}$. The spin relaxation time is expressed as $1/T_1^{\mathrm{spin}}=A\Delta_{\mathrm{gs}}^3(n_{\mathrm{BE}}(\Delta_{\mathrm{gs}},T))$ \cite{Jahnke2015}, which results in a further  drastic increase in the relaxation time at $k_\textrm{B}T<\Delta_{\mathrm{gs}}$.

\begin{figure}
	\centering
		\includegraphics[width=80 mm]{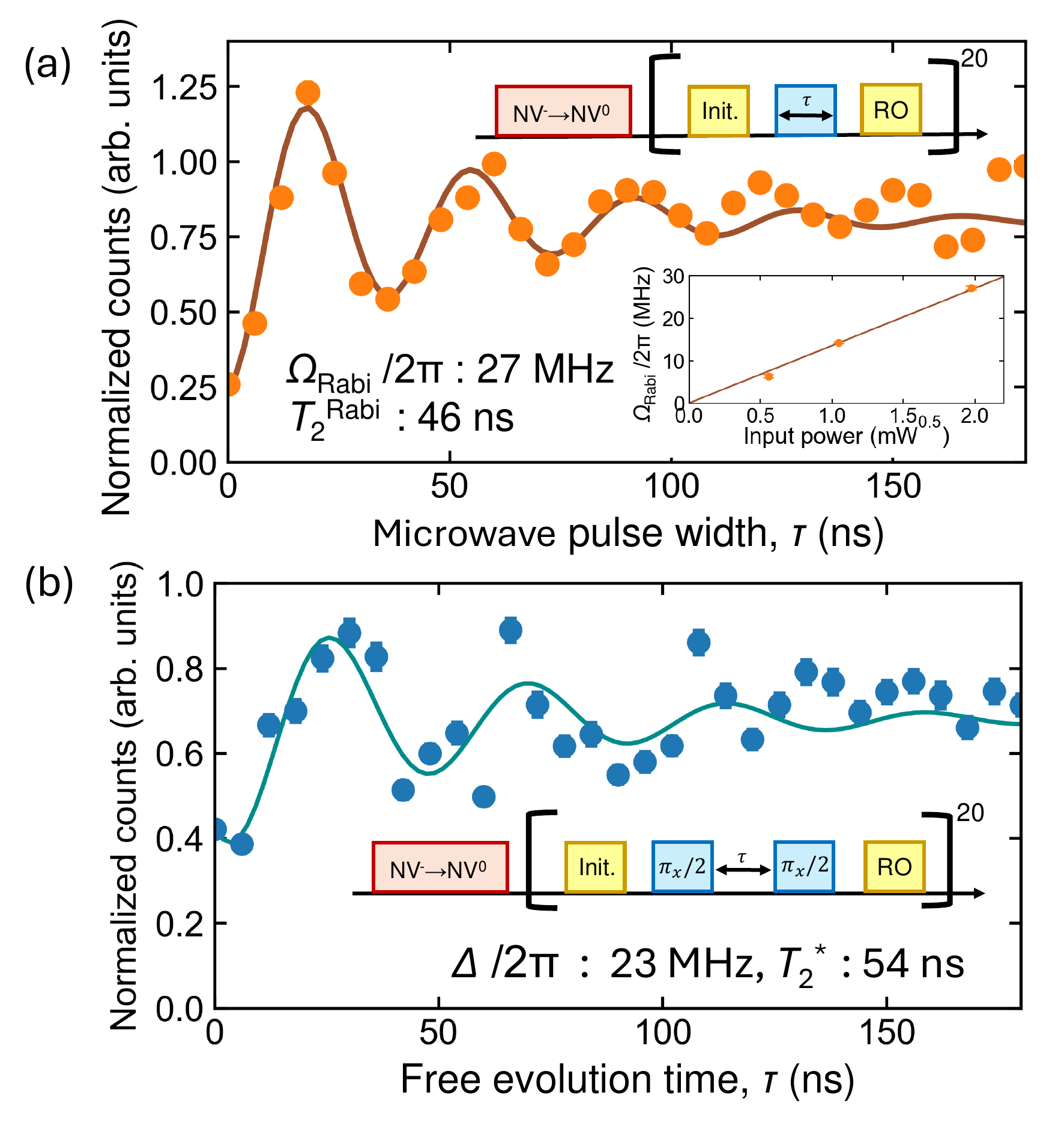}
	\caption{The Rabi oscillation and the Ramsey interference with their measurement sequences. (a) The Rabi oscillation between the orbital states, $\ket{0}$ and $\ket{1}$. The driving power outside the dilution refrigerator is 3.9 mW. The data are fitted with a function, $A \mathrm{cos}(\varOmega_{\mathrm{Rabi}} t)\mathrm{exp}(-t/T_2^{\mathrm{Rabi}})+ B$ , where $A$ is the amplitude, $B$ is the center of the oscillation, $\varOmega_\mathrm{Rabi}$ is the Rabi frequency. The inset shows the Rabi frequency as a function of the square root of the driving power. The error bars correspond to the standard deviation considering the photon shot noise. (b) Ramsey interference when the driving frequency is set to be 13,020 MHz. The fitting function is: $C \mathrm{cos}(\Delta t +\phi) \mathrm{exp}(-t/T_2^*)+ D$, where $C$ is the amplitude, $\Delta$ is the detuning from the resonant frequency, $\phi$ is the phase offset, $D$ is the center of the oscillation. The error bars correspond to the standard deviation considering the photon shot noise.}
	\label{fig:Rabi}
\end{figure}

%%Rabi Ramsey
After characterizing the orbital relaxation time, we proceed to perform coherent control of the orbital states and estimate the orbital coherence time. Following the orbital initialization from $\ket{0}$ to $\ket{1}$ using a 575-nm pulse, a microwave pulse resonant with  $\Delta_{\textrm{gs}}$ is applied with changing the width. The time delay of the readout pulse is also adjusted as the pulse width increases. Rabi oscillation between $\ket{0}$ and $\ket{1}$ with a decaying time constant ($T_2^{\mathrm{Rabi}}$) of 46$\pm2$ ns is observed as shown in Fig. \ref{fig:Rabi} (a). Maximum Rabi frequency is 27 MHz with an input peak power of 3.9 mW, which is measured at outside the dilution refrigerator. The Rabi frequency is proportional to the square root of the input power as shown in the inset of Fig. \ref{fig:Rabi} (a), indicating that the detuning is negligibly small. Throughout all measurements conducted at various input powers, the temperature at the mixing chamber plate is maintained at 50 mK. The maximum average input power at a microwave port outside the dilution refrigerator is 41 $\si[per-mode=symbol]{\micro \watt}$ (See Supplementary Notes for the temperature increase of the sample during the measurements). %Since there is a distance between the sample and a thermometer, we estimate the increase of local temperature during the measurements based on the change in $T_1^\textrm{orb}$ by the application of microwave pulse. The increase of local temperature is estimated to be $\sim$400 mK when the average input microwave power outside the dilution refrigerator is 1.6 $\mu$W which corresponds to 3.9 mW input with 150 ns . 

Additionally, we estimate the initialization fidelity as follows. Assuming that the quantum state reaches the complete mixed state after the application of a microwave driving pulse much longer than $T_2^{\textrm{Rabi}}$, the normalized counts for the $\ket{0}$ state are expected to be twice the normalized counts after the 180-ns pulse, $(0.8\pm0.02)\times2=1.6\pm0.04$. Since the normalized counts at 0 ns pulse width is 0.24, the initialization fidelity is estimated to be 85$\pm3\%$ (calculated as $1-0.24/1.6$) for a 1-$\si[per-mode=symbol]{\micro \second}$ orbital initialization pulse used in these experiments. The initialization fidelity is enhanced up to around 93$\%$ (See Supplementary Notes) by increasing the initialization pulse width from 1 $\si[per-mode=symbol]{\micro \second}$ to 3 $\si[per-mode=symbol]{\micro \second}$, which is slightly lower than that the estimated value ($\sim97\%$) based on the optical-Bloch equation (See Supplementary Notes). The above estimate also indicates that the populations in $\ket{0}$ and $\ket{1}$ before initialization are 0.625 and 0.375, respectively. The occupation of $\ket{1}$ is higher than that expected from the Fermi-Dirac distribution for an electron of NV$^0$ and the Bose-Einstein distribution for phonons in the environment at 50 mK, which is attributed to microwave induced heating (See Supplementary Notes).

The orbital coherence time, $T_2^*$, is estimated to be 54$\pm$5 ns from the Ramsey interference measurement (Fig. \ref{fig:Rabi}(b)) with a certain detuning from $\Delta_{\mathrm{gs}}$. The value of $T_2^*$ is close to that of $T_2^{\mathrm{Rabi}}$, indicating that similar noise processes can dominate $T_2^*$ and $T_2^{\mathrm{Rabi}}$. According to Ref. \cite{Kurokawa2024}, $T_2^*$ at 5 K is 31$\pm$3.6 ns. In contrast to the orbital relaxation time, the orbital coherence time does not increase with decreasing temperature, indicating it is not limited by the thermal phonon in the environment.

\begin{figure}
	\centering
		\includegraphics[width=80 mm]{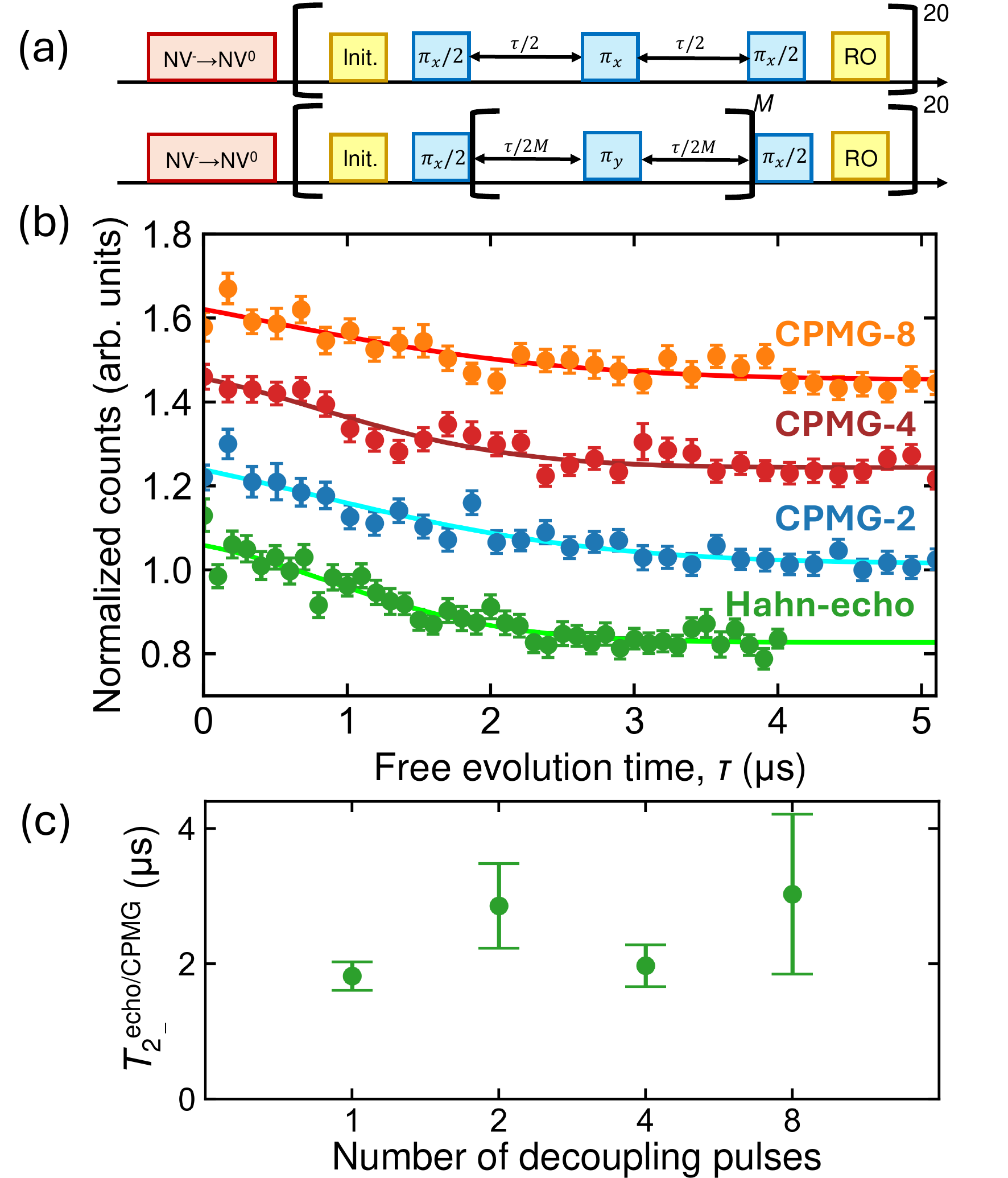}
	\caption{Coherence extension by dynamical decoupling. (a) The pulse sequences for Hahn-echo (top) and Carr-Purcell-Meiboom-Gill (CPMG)-$M$ (bottom). The lowercase notation $\pi_{x(y)}$ indicates that the phase of bit-flipping pulse is in(quadrature)-phase. (b) Coherence decay under the application of dynamical decoupling pulses (Hahn-echo, CPMG-$M$ ($M$ = 2, 4, 8)), with offsets applied to each curve. The decaying curves are fitted with the function  $E\mathrm{exp}(-(t/T_2^{\textrm{echo/CPMG}})^2-t/T_1^{\textrm{orb}}) + F$, where $E$ and $F$, and $n$ are some constants. The error bars correspond to the standard deviation considering the photon shot noise. (c) The number of the decoupling pulses versus the extended coherence time. The error bars correspond to the standard error in the nonlinear fitting.
	}
	\label{fig:echo}
\end{figure}

%%echo
We also apply the dynamical decoupling techniques to extend the orbital coherence time of NV$^0$. By applying bit-flipping pulses (Fig. \ref{fig:echo} (a)), the dynamically decoupled coherence time, $T_2^{\textrm{echo/CPMG}}$, increases to 1.8$\pm$0.2 $\si[per-mode=symbol]{\micro \second}$ for Hahn-echo and 2-3 $\si[per-mode=symbol]{\micro \second}$ for the  Carr-Purcell-Meiboom-Gill (CPMG) sequence (Fig. \ref{fig:echo} (b)(c)). A more than 30-fold increase in the orbital coherence time is achieved by applying these decoupling techniques. The dynamical decoupling sequences filter out low-frequency phase noise during experiments \cite{Krantz2021}. Considering that a decrease in temperature does not contribute to the increase in orbital coherence time, $T_2^*$ is mainly limited by spectral diffusion during the measurements, which occurs on a relatively slow time scale compared to $T_2^{\textrm{echo/CPMG}}$.  The enhancement in $T_2^{\textrm{echo/CPMG}}$ is modest with increasing in the number of decoupling pulses (Fig. \ref{fig:echo} (b)(c)), which is attributed to the decrease in $T_1^\mathrm{orb}$ down to 1.3-1.8 $\si[per-mode=symbol]{\micro \second}$ due to the local heating from the driving microwave (See Supplementary Notes). %$T_2^{\textrm{echo/CPMG}}$ is mainly limited by $T_1^\mathrm{orb}$. Additionally, the noise process limiting $T_2^{\textrm{echo/CPMG}}$ is likely a Markovian white noise.

%Since increasing the number of pulses shifts the noise filter function to higher frequencies \cite{Krantz2021}, the noise process limiting $T_2^{\textrm{echo/CPMG}}$ can be a Markovian white noise.

%, 2.1$\pm$4.9 $\mu$s for CPMG-2, 1.7$\pm$0.3 $\mu$s for CPMG-4, and 2.1$\pm$0.6 $\mu$s for CPMG-8

%%simulaiton?

Based on the relaxation and coherence times obtained from the above measurements, we explore the possibility of realizing cavity quantum electrodynamics using NV$^0$ and a microwave photon. The single microwave photon voltage generated by the state-of-the-art microwave resonator with 4-k$\Omega$ impedance \cite{Samkharadze2016,Niepce2019a} is $\omega_\mathrm{mw}\sqrt{\hbar Z/2}\sim$ 37 $\si[per-mode=symbol]{\micro \volt}$, where $\omega_\mathrm{mw}/2\pi$ = 13 GHz is the angular frequency of the microwave resonator, $Z$ is the characteristic impedance of the microwave resonator. Considering that NV$^0$ is placed between the gap of a ring microwave resonator, the electric field between the gap is $37\times2/d$ $\si[per-mode=symbol]{\micro \volt \per \meter}$, where $d$ is the distance between the gap. The factor of 2 is introduced because one end of the ring resonator has +37 $\si[per-mode=symbol]{\micro \volt}$ and the other end has -37 $\si[per-mode=symbol]{\micro \volt}$. When we set $d$ to be 1 $\si[per-mode=symbol]{\micro \meter}$, the resulting electric field is 0.74 V/cm. Since the electric field sensitivity of NV$^0$ is 1 MHz/V cm$^{-1}$ \cite{Kurokawa2024}, the coupling strength, $g_{\textrm{mw-NV0}}/2\pi$, between the microwave electric field and NV$^0$ is 740 kHz. Assuming the relaxation rate of NV$^0$, $\gamma_1^{\textrm{NV0}}/2\pi$, to be 34 kHz (=$1/2\pi T_1^{\textrm{orb}}$), and that of the microwave resonator, $\kappa_{\textrm{mw}}/2\pi$, to be 100 kHz ($Q_\textrm{mw}$ = 13 GHz/100 kHz=$1.3\times10^5$), the strong coupling regime, in which  $g_\textrm{mw-NV0}>\gamma_1^{\textrm{NV0}},\kappa_{\textrm{mw}}$ and $C=4g_{\textrm{mw-NV0}}^2/\gamma_1^{\textrm{NV0}}\kappa_{\textrm{mw}}=664$, is achievable, where $C$ is the cooperativity. Even stronger coupling can be achieved when NV$^0$ is placed in a phononic crystal cavity. Assuming that the strain sensitivity of NV$^0$ is comparable to the excited state of NV$^-$ \cite{Venturi2019}, the coupling between NV$^0$ and a phonon mode can be 1-10 MHz \cite{Kim2023b} due to the small mode volume of the phononic crystal. However, we must overcome the spectral diffusion or hopping of  NV$^0$ to ensure a stable experiment. In the presence of the spectral hopping, $T_2^*$ can limit the coherent interaction time. Additionally, the integration into a nanostructure like a phononic crystal, presents challenges, as it can induce significant spectral diffusion for NV centers.

%\section*{Discussion}
In conclusion, we have demonstrated quantum control of NV$^0$ in a dilution refrigerator and have measured the orbital relaxation and coherence times. The orbital relaxation time shows an approximately tenfold increase compared to that at 5 K. The orbital coherence time have not exhibited an improvement just by decreasing the temperature. However, we have extended the  coherence time to around 1.8 $\si[per-mode=symbol]{\micro \second}$ using the dynamical decoupling. This is a 30-fold increase compared to the coherence time without the dynamical decoupling. Based on these results, the strong coupling regime between a single  NV$^0$ and a high-impedance microwave resonator or a phononic crystal cavity is within reach. Compared to group IV color centers, the relatively small $\Delta_{\mathrm{gs}}$ ($\sim$10 GHz) and high electric field sensitivity of NV$^0$ are advantages realizing strong coupling to other fields. Although the stabilization of the spectrum remains a challenge, this work opens the possibility of using the optically active defect center, NV$^0$, as a quantum interface between different degrees of freedom: microwave photon, microwave phonon, and optical photon.%, and also an optical photon.

\section*{Acknowledgements}
H. Kosaka acknowledges the funding support from Japan Science and Technology Agency (JST) Moonshot R$\&$D grant (JPMJMS2062) and JST CREST grant (JPMJCR1773). H. Kosaka also acknowledges the Ministry of Internal Affairs and Communications (MIC) for funding, research and development for construction of global quantum cryptography network (JPMI00316), and the Japan Society for the Promotion of Science (JSPS) Grants-in-Aid for Scientific Research (20H05661, 20K20441).
%This work was supported by JSPS KAKENHI Grant Numbers  JP20H05164, JP19K14661.
%%
\section*{Authorship contribution}
H. Kurokawa designed and performed the experiments. S. Nakazato also performed experiments in a different sample. T. Makino and H. Kato fabricated the electrical circuit. S. Onoda contributed to the creation of NV centers by the electron beam irradiation and annealing. Y. Sekiguchi contributed the fabrication of the solid-immersion-lens structure. H. Kosaka supervised the project.

\section*{Competing interests}
The authors declare no competing interests.


%apsrev4-2.bst 2019-01-14 (MD) hand-edited version of apsrev4-1.bst
%Control: key (0)
%Control: author (72) initials jnrlst
%Control: editor formatted (1) identically to author
%Control: production of article title (-1) disabled
%Control: page (0) single
%Control: year (1) truncated
%Control: production of eprint (0) enabled
\begin{thebibliography}{2}%
\makeatletter
\providecommand \@ifxundefined [1]{%
 \@ifx{#1\undefined}
}%
\providecommand \@ifnum [1]{%
 \ifnum #1\expandafter \@firstoftwo
 \else \expandafter \@secondoftwo
 \fi
}%
\providecommand \@ifx [1]{%
 \ifx #1\expandafter \@firstoftwo
 \else \expandafter \@secondoftwo
 \fi
}%
\providecommand \natexlab [1]{#1}%
\providecommand \enquote  [1]{``#1''}%
\providecommand \bibnamefont  [1]{#1}%
\providecommand \bibfnamefont [1]{#1}%
\providecommand \citenamefont [1]{#1}%
\providecommand \href@noop [0]{\@secondoftwo}%
\providecommand \href [0]{\begingroup \@sanitize@url \@href}%
\providecommand \@href[1]{\@@startlink{#1}\@@href}%
\providecommand \@@href[1]{\endgroup#1\@@endlink}%
\providecommand \@sanitize@url [0]{\catcode `\\12\catcode `\$12\catcode
  `\&12\catcode `\#12\catcode `\^12\catcode `\_12\catcode `\%12\relax}%
\providecommand \@@startlink[1]{}%
\providecommand \@@endlink[0]{}%
\providecommand \url  [0]{\begingroup\@sanitize@url \@url }%
\providecommand \@url [1]{\endgroup\@href {#1}{\urlprefix }}%
\providecommand \urlprefix  [0]{URL }%
\providecommand \Eprint [0]{\href }%
\providecommand \doibase [0]{https://doi.org/}%
\providecommand \selectlanguage [0]{\@gobble}%
\providecommand \bibinfo  [0]{\@secondoftwo}%
\providecommand \bibfield  [0]{\@secondoftwo}%
\providecommand \translation [1]{[#1]}%
\providecommand \BibitemOpen [0]{}%
\providecommand \bibitemStop [0]{}%
\providecommand \bibitemNoStop [0]{.\EOS\space}%
\providecommand \EOS [0]{\spacefactor3000\relax}%
\providecommand \BibitemShut  [1]{\csname bibitem#1\endcsname}%
\let\auto@bib@innerbib\@empty
%</preamble>
\bibitem [{\citenamefont {Baier}\ \emph {et~al.}(2020)\citenamefont {Baier},
  \citenamefont {Bradley}, \citenamefont {Middelburg}, \citenamefont
  {Dobrovitski}, \citenamefont {Taminiau},\ and\ \citenamefont
  {Hanson}}]{Baier2020}%
  \BibitemOpen
  \bibfield  {author} {\bibinfo {author} {\bibfnamefont {S.}~\bibnamefont
  {Baier}}, \bibinfo {author} {\bibfnamefont {C.~E.}\ \bibnamefont {Bradley}},
  \bibinfo {author} {\bibfnamefont {T.}~\bibnamefont {Middelburg}}, \bibinfo
  {author} {\bibfnamefont {V.~V.}\ \bibnamefont {Dobrovitski}}, \bibinfo
  {author} {\bibfnamefont {T.~H.}\ \bibnamefont {Taminiau}},\ and\ \bibinfo
  {author} {\bibfnamefont {R.}~\bibnamefont {Hanson}},\ }\href
  {https://doi.org/10.1103/PhysRevLett.125.193601} {\bibfield  {journal}
  {\bibinfo  {journal} {Physical Review Letters}\ }\textbf {\bibinfo {volume}
  {125}},\ \bibinfo {pages} {193601} (\bibinfo {year} {2020})},\ \Eprint
  {https://arxiv.org/abs/2007.14673} {arXiv:2007.14673} \BibitemShut {NoStop}%
\bibitem [{\citenamefont {Jahnke}\ \emph {et~al.}(2015)\citenamefont {Jahnke},
  \citenamefont {Sipahigil}, \citenamefont {Binder}, \citenamefont {Doherty},
  \citenamefont {Metsch}, \citenamefont {Rogers}, \citenamefont {Manson},
  \citenamefont {Lukin},\ and\ \citenamefont {Jelezko}}]{Jahnke2015}%
  \BibitemOpen
  \bibfield  {author} {\bibinfo {author} {\bibfnamefont {K.~D.}\ \bibnamefont
  {Jahnke}}, \bibinfo {author} {\bibfnamefont {A.}~\bibnamefont {Sipahigil}},
  \bibinfo {author} {\bibfnamefont {J.~M.}\ \bibnamefont {Binder}}, \bibinfo
  {author} {\bibfnamefont {M.~W.}\ \bibnamefont {Doherty}}, \bibinfo {author}
  {\bibfnamefont {M.}~\bibnamefont {Metsch}}, \bibinfo {author} {\bibfnamefont
  {L.~J.}\ \bibnamefont {Rogers}}, \bibinfo {author} {\bibfnamefont {N.~B.}\
  \bibnamefont {Manson}}, \bibinfo {author} {\bibfnamefont {M.~D.}\
  \bibnamefont {Lukin}},\ and\ \bibinfo {author} {\bibfnamefont
  {F.}~\bibnamefont {Jelezko}},\ }\href
  {https://doi.org/10.1088/1367-2630/17/4/043011} {\bibfield  {journal}
  {\bibinfo  {journal} {New Journal of Physics}\ }\textbf {\bibinfo {volume}
  {17}},\ \bibinfo {pages} {043011} (\bibinfo {year} {2015})},\ \Eprint
  {https://arxiv.org/abs/1411.2871} {arXiv:1411.2871} \BibitemShut {NoStop}%
\end{thebibliography}%


\begin{thebibliography}{42}%
\makeatletter
\providecommand \@ifxundefined [1]{%
 \@ifx{#1\undefined}
}%
\providecommand \@ifnum [1]{%
 \ifnum #1\expandafter \@firstoftwo
 \else \expandafter \@secondoftwo
 \fi
}%
\providecommand \@ifx [1]{%
 \ifx #1\expandafter \@firstoftwo
 \else \expandafter \@secondoftwo
 \fi
}%
\providecommand \natexlab [1]{#1}%
\providecommand \enquote  [1]{``#1''}%
\providecommand \bibnamefont  [1]{#1}%
\providecommand \bibfnamefont [1]{#1}%
\providecommand \citenamefont [1]{#1}%
\providecommand \href@noop [0]{\@secondoftwo}%
\providecommand \href [0]{\begingroup \@sanitize@url \@href}%
\providecommand \@href[1]{\@@startlink{#1}\@@href}%
\providecommand \@@href[1]{\endgroup#1\@@endlink}%
\providecommand \@sanitize@url [0]{\catcode `\\12\catcode `\$12\catcode
  `\&12\catcode `\#12\catcode `\^12\catcode `\_12\catcode `\%12\relax}%
\providecommand \@@startlink[1]{}%
\providecommand \@@endlink[0]{}%
\providecommand \url  [0]{\begingroup\@sanitize@url \@url }%
\providecommand \@url [1]{\endgroup\@href {#1}{\urlprefix }}%
\providecommand \urlprefix  [0]{URL }%
\providecommand \Eprint [0]{\href }%
\providecommand \doibase [0]{https://doi.org/}%
\providecommand \selectlanguage [0]{\@gobble}%
\providecommand \bibinfo  [0]{\@secondoftwo}%
\providecommand \bibfield  [0]{\@secondoftwo}%
\providecommand \translation [1]{[#1]}%
\providecommand \BibitemOpen [0]{}%
\providecommand \bibitemStop [0]{}%
\providecommand \bibitemNoStop [0]{.\EOS\space}%
\providecommand \EOS [0]{\spacefactor3000\relax}%
\providecommand \BibitemShut  [1]{\csname bibitem#1\endcsname}%
\let\auto@bib@innerbib\@empty
%</preamble>
\bibitem [{\citenamefont {Schirhagl}\ \emph {et~al.}(2014)\citenamefont
  {Schirhagl}, \citenamefont {Chang}, \citenamefont {Loretz},\ and\
  \citenamefont {Degen}}]{Schirhagl2014}%
  \BibitemOpen
  \bibfield  {author} {\bibinfo {author} {\bibfnamefont {R.}~\bibnamefont
  {Schirhagl}}, \bibinfo {author} {\bibfnamefont {K.}~\bibnamefont {Chang}},
  \bibinfo {author} {\bibfnamefont {M.}~\bibnamefont {Loretz}},\ and\ \bibinfo
  {author} {\bibfnamefont {C.~L.}\ \bibnamefont {Degen}},\ }\bibfield  {title}
  {\bibinfo {title} {{Nitrogen-Vacancy Centers in Diamond : Nanoscale Sensors
  for Physics and Biology}},\ }\href
  {https://doi.org/10.1146/annurev-physchem-040513-103659} {\bibfield
  {journal} {\bibinfo  {journal} {Annu. Rev. Phys. Chem}\ }\textbf {\bibinfo
  {volume} {65}},\ \bibinfo {pages} {83} (\bibinfo {year} {2014})}\BibitemShut
  {NoStop}%
\bibitem [{\citenamefont {Beveratos}\ \emph {et~al.}(2002)\citenamefont
  {Beveratos}, \citenamefont {K{\"{u}}hn}, \citenamefont {Brouri},
  \citenamefont {Gacoin}, \citenamefont {Poizat},\ and\ \citenamefont
  {Grangier}}]{Beveratos2002}%
  \BibitemOpen
  \bibfield  {author} {\bibinfo {author} {\bibfnamefont {A.}~\bibnamefont
  {Beveratos}}, \bibinfo {author} {\bibfnamefont {S.}~\bibnamefont
  {K{\"{u}}hn}}, \bibinfo {author} {\bibfnamefont {R.}~\bibnamefont {Brouri}},
  \bibinfo {author} {\bibfnamefont {T.}~\bibnamefont {Gacoin}}, \bibinfo
  {author} {\bibfnamefont {J.~P.}\ \bibnamefont {Poizat}},\ and\ \bibinfo
  {author} {\bibfnamefont {P.}~\bibnamefont {Grangier}},\ }\bibfield  {title}
  {\bibinfo {title} {{Room temperature stable single-photon source}},\ }\href
  {https://doi.org/10.1140/epjd/e20020023} {\bibfield  {journal} {\bibinfo
  {journal} {European Physical Journal D}\ }\textbf {\bibinfo {volume} {18}},\
  \bibinfo {pages} {191} (\bibinfo {year} {2002})}\BibitemShut {NoStop}%
\bibitem [{\citenamefont {Babinec}\ \emph {et~al.}(2010)\citenamefont
  {Babinec}, \citenamefont {Hausmann}, \citenamefont {Khan}, \citenamefont
  {Zhang}, \citenamefont {Maze}, \citenamefont {Hemmer},\ and\ \citenamefont
  {Lon{\v{c}}ar}}]{Babinec2010}%
  \BibitemOpen
  \bibfield  {author} {\bibinfo {author} {\bibfnamefont {T.~M.}\ \bibnamefont
  {Babinec}}, \bibinfo {author} {\bibfnamefont {B.~J.~M.}\ \bibnamefont
  {Hausmann}}, \bibinfo {author} {\bibfnamefont {M.}~\bibnamefont {Khan}},
  \bibinfo {author} {\bibfnamefont {Y.}~\bibnamefont {Zhang}}, \bibinfo
  {author} {\bibfnamefont {J.~R.}\ \bibnamefont {Maze}}, \bibinfo {author}
  {\bibfnamefont {P.~R.}\ \bibnamefont {Hemmer}},\ and\ \bibinfo {author}
  {\bibfnamefont {M.}~\bibnamefont {Lon{\v{c}}ar}},\ }\bibfield  {title}
  {\bibinfo {title} {{A diamond nanowire single-photon source}},\ }\href
  {https://doi.org/10.1038/nnano.2010.6} {\bibfield  {journal} {\bibinfo
  {journal} {Nature Nanotechnology}\ }\textbf {\bibinfo {volume} {5}},\
  \bibinfo {pages} {195} (\bibinfo {year} {2010})}\BibitemShut {NoStop}%
\bibitem [{\citenamefont {Mizuochi}\ \emph {et~al.}(2012)\citenamefont
  {Mizuochi}, \citenamefont {Makino}, \citenamefont {Kato}, \citenamefont
  {Takeuchi}, \citenamefont {Ogura}, \citenamefont {Okushi}, \citenamefont
  {Nothaft}, \citenamefont {Neumann}, \citenamefont {Gali}, \citenamefont
  {Jelezko}, \citenamefont {Wrachtrup},\ and\ \citenamefont
  {Yamasaki}}]{Mizuochi2012}%
  \BibitemOpen
  \bibfield  {author} {\bibinfo {author} {\bibfnamefont {N.}~\bibnamefont
  {Mizuochi}}, \bibinfo {author} {\bibfnamefont {T.}~\bibnamefont {Makino}},
  \bibinfo {author} {\bibfnamefont {H.}~\bibnamefont {Kato}}, \bibinfo {author}
  {\bibfnamefont {D.}~\bibnamefont {Takeuchi}}, \bibinfo {author}
  {\bibfnamefont {M.}~\bibnamefont {Ogura}}, \bibinfo {author} {\bibfnamefont
  {H.}~\bibnamefont {Okushi}}, \bibinfo {author} {\bibfnamefont
  {M.}~\bibnamefont {Nothaft}}, \bibinfo {author} {\bibfnamefont
  {P.}~\bibnamefont {Neumann}}, \bibinfo {author} {\bibfnamefont
  {A.}~\bibnamefont {Gali}}, \bibinfo {author} {\bibfnamefont {F.}~\bibnamefont
  {Jelezko}}, \bibinfo {author} {\bibfnamefont {J.}~\bibnamefont {Wrachtrup}},\
  and\ \bibinfo {author} {\bibfnamefont {S.}~\bibnamefont {Yamasaki}},\
  }\bibfield  {title} {\bibinfo {title} {{Electrically driven single-photon
  source at room temperature in diamond}},\ }\href
  {https://doi.org/10.1038/nphoton.2012.75} {\bibfield  {journal} {\bibinfo
  {journal} {Nature Photonics}\ }\textbf {\bibinfo {volume} {6}},\ \bibinfo
  {pages} {299} (\bibinfo {year} {2012})}\BibitemShut {NoStop}%
\bibitem [{\citenamefont {Knall}\ \emph {et~al.}(2022)\citenamefont {Knall},
  \citenamefont {Knaut}, \citenamefont {Bekenstein}, \citenamefont {Assumpcao},
  \citenamefont {Stroganov}, \citenamefont {Gong}, \citenamefont {Huan},
  \citenamefont {Stas}, \citenamefont {Machielse}, \citenamefont {Chalupnik},
  \citenamefont {Levonian}, \citenamefont {Suleymanzade}, \citenamefont
  {Riedinger}, \citenamefont {Park}, \citenamefont {Lon{\v{c}}ar},
  \citenamefont {Bhaskar},\ and\ \citenamefont {Lukin}}]{Knall2022}%
  \BibitemOpen
  \bibfield  {author} {\bibinfo {author} {\bibfnamefont {E.~N.}\ \bibnamefont
  {Knall}}, \bibinfo {author} {\bibfnamefont {C.~M.}\ \bibnamefont {Knaut}},
  \bibinfo {author} {\bibfnamefont {R.}~\bibnamefont {Bekenstein}}, \bibinfo
  {author} {\bibfnamefont {D.~R.}\ \bibnamefont {Assumpcao}}, \bibinfo {author}
  {\bibfnamefont {P.~L.}\ \bibnamefont {Stroganov}}, \bibinfo {author}
  {\bibfnamefont {W.}~\bibnamefont {Gong}}, \bibinfo {author} {\bibfnamefont
  {Y.~Q.}\ \bibnamefont {Huan}}, \bibinfo {author} {\bibfnamefont {P.-J.}\
  \bibnamefont {Stas}}, \bibinfo {author} {\bibfnamefont {B.}~\bibnamefont
  {Machielse}}, \bibinfo {author} {\bibfnamefont {M.}~\bibnamefont
  {Chalupnik}}, \bibinfo {author} {\bibfnamefont {D.}~\bibnamefont {Levonian}},
  \bibinfo {author} {\bibfnamefont {A.}~\bibnamefont {Suleymanzade}}, \bibinfo
  {author} {\bibfnamefont {R.}~\bibnamefont {Riedinger}}, \bibinfo {author}
  {\bibfnamefont {H.}~\bibnamefont {Park}}, \bibinfo {author} {\bibfnamefont
  {M.}~\bibnamefont {Lon{\v{c}}ar}}, \bibinfo {author} {\bibfnamefont {M.~K.}\
  \bibnamefont {Bhaskar}},\ and\ \bibinfo {author} {\bibfnamefont {M.~D.}\
  \bibnamefont {Lukin}},\ }\bibfield  {title} {\bibinfo {title} {{Efficient
  Source of Shaped Single Photons Based on an Integrated Diamond Nanophotonic
  System}},\ }\href {https://doi.org/10.1103/PhysRevLett.129.053603} {\bibfield
   {journal} {\bibinfo  {journal} {Physical Review Letters}\ }\textbf {\bibinfo
  {volume} {129}},\ \bibinfo {pages} {053603} (\bibinfo {year}
  {2022})}\BibitemShut {NoStop}%
\bibitem [{\citenamefont {Nemoto}\ \emph {et~al.}(2014)\citenamefont {Nemoto},
  \citenamefont {Trupke}, \citenamefont {Devitt}, \citenamefont {Stephens},
  \citenamefont {Scharfenberger}, \citenamefont {Buczak}, \citenamefont
  {N{\"{o}}bauer}, \citenamefont {Everitt}, \citenamefont {Schmiedmayer},\ and\
  \citenamefont {Munro}}]{Nemoto2014}%
  \BibitemOpen
  \bibfield  {author} {\bibinfo {author} {\bibfnamefont {K.}~\bibnamefont
  {Nemoto}}, \bibinfo {author} {\bibfnamefont {M.}~\bibnamefont {Trupke}},
  \bibinfo {author} {\bibfnamefont {S.~J.}\ \bibnamefont {Devitt}}, \bibinfo
  {author} {\bibfnamefont {A.~M.}\ \bibnamefont {Stephens}}, \bibinfo {author}
  {\bibfnamefont {B.}~\bibnamefont {Scharfenberger}}, \bibinfo {author}
  {\bibfnamefont {K.}~\bibnamefont {Buczak}}, \bibinfo {author} {\bibfnamefont
  {T.}~\bibnamefont {N{\"{o}}bauer}}, \bibinfo {author} {\bibfnamefont {M.~S.}\
  \bibnamefont {Everitt}}, \bibinfo {author} {\bibfnamefont {J.}~\bibnamefont
  {Schmiedmayer}},\ and\ \bibinfo {author} {\bibfnamefont {W.~J.}\ \bibnamefont
  {Munro}},\ }\bibfield  {title} {\bibinfo {title} {{Photonic Architecture for
  Scalable Quantum Information Processing in Diamond}},\ }\href
  {https://doi.org/10.1103/PhysRevX.4.031022} {\bibfield  {journal} {\bibinfo
  {journal} {Physical Review X}\ }\textbf {\bibinfo {volume} {4}},\ \bibinfo
  {pages} {031022} (\bibinfo {year} {2014})}\BibitemShut {NoStop}%
\bibitem [{\citenamefont {Pezzagna}\ and\ \citenamefont
  {Meijer}(2021)}]{Pezzagna2021}%
  \BibitemOpen
  \bibfield  {author} {\bibinfo {author} {\bibfnamefont {S.}~\bibnamefont
  {Pezzagna}}\ and\ \bibinfo {author} {\bibfnamefont {J.}~\bibnamefont
  {Meijer}},\ }\bibfield  {title} {\bibinfo {title} {{Quantum computer based on
  color centers in diamond}},\ }\href
  {https://doi.org/https://doi.org/10.1063/5.0007444} {\bibfield  {journal}
  {\bibinfo  {journal} {Applied Physics Reviews}\ }\textbf {\bibinfo {volume}
  {8}},\ \bibinfo {pages} {011308} (\bibinfo {year} {2021})}\BibitemShut
  {NoStop}%
\bibitem [{\citenamefont {Abobeih}\ \emph {et~al.}(2022)\citenamefont
  {Abobeih}, \citenamefont {Wang}, \citenamefont {Randall}, \citenamefont
  {Loenen}, \citenamefont {Bradley}, \citenamefont {Markham}, \citenamefont
  {Twitchen}, \citenamefont {Terhal},\ and\ \citenamefont
  {Taminiau}}]{Abobeih2022}%
  \BibitemOpen
  \bibfield  {author} {\bibinfo {author} {\bibfnamefont {M.~H.}\ \bibnamefont
  {Abobeih}}, \bibinfo {author} {\bibfnamefont {Y.}~\bibnamefont {Wang}},
  \bibinfo {author} {\bibfnamefont {J.}~\bibnamefont {Randall}}, \bibinfo
  {author} {\bibfnamefont {S.~J.~H.}\ \bibnamefont {Loenen}}, \bibinfo {author}
  {\bibfnamefont {C.~E.}\ \bibnamefont {Bradley}}, \bibinfo {author}
  {\bibfnamefont {M.}~\bibnamefont {Markham}}, \bibinfo {author} {\bibfnamefont
  {D.~J.}\ \bibnamefont {Twitchen}}, \bibinfo {author} {\bibfnamefont {B.~M.}\
  \bibnamefont {Terhal}},\ and\ \bibinfo {author} {\bibfnamefont {T.~H.}\
  \bibnamefont {Taminiau}},\ }\bibfield  {title} {\bibinfo {title}
  {{Fault-tolerant operation of a logical qubit in a diamond quantum
  processor}},\ }\href {https://doi.org/10.1038/s41586-022-04819-6} {\bibfield
  {journal} {\bibinfo  {journal} {Nature}\ }\textbf {\bibinfo {volume} {606}},\
  \bibinfo {pages} {884} (\bibinfo {year} {2022})}\BibitemShut {NoStop}%
\bibitem [{\citenamefont {Sekiguchi}\ \emph {et~al.}(2022)\citenamefont
  {Sekiguchi}, \citenamefont {Matsushita}, \citenamefont {Kawasaki},\ and\
  \citenamefont {Kosaka}}]{Sekiguchi2022}%
  \BibitemOpen
  \bibfield  {author} {\bibinfo {author} {\bibfnamefont {Y.}~\bibnamefont
  {Sekiguchi}}, \bibinfo {author} {\bibfnamefont {K.}~\bibnamefont
  {Matsushita}}, \bibinfo {author} {\bibfnamefont {Y.}~\bibnamefont
  {Kawasaki}},\ and\ \bibinfo {author} {\bibfnamefont {H.}~\bibnamefont
  {Kosaka}},\ }\bibfield  {title} {\bibinfo {title} {{Optically addressable
  universal holonomic quantum gates on diamond spins}},\ }\href
  {https://doi.org/10.1038/s41566-022-01038-3} {\bibfield  {journal} {\bibinfo
  {journal} {Nature Photonics}\ }\textbf {\bibinfo {volume} {16}},\ \bibinfo
  {pages} {662} (\bibinfo {year} {2022})}\BibitemShut {NoStop}%
\bibitem [{\citenamefont {Sekiguchi}\ \emph {et~al.}(2017)\citenamefont
  {Sekiguchi}, \citenamefont {Niikura}, \citenamefont {Kuroiwa}, \citenamefont
  {Kano},\ and\ \citenamefont {Kosaka}}]{Sekiguchi2017}%
  \BibitemOpen
  \bibfield  {author} {\bibinfo {author} {\bibfnamefont {Y.}~\bibnamefont
  {Sekiguchi}}, \bibinfo {author} {\bibfnamefont {N.}~\bibnamefont {Niikura}},
  \bibinfo {author} {\bibfnamefont {R.}~\bibnamefont {Kuroiwa}}, \bibinfo
  {author} {\bibfnamefont {H.}~\bibnamefont {Kano}},\ and\ \bibinfo {author}
  {\bibfnamefont {H.}~\bibnamefont {Kosaka}},\ }\bibfield  {title} {\bibinfo
  {title} {{Optical holonomic single quantum gates with a geometric spin under
  a zero field}},\ }\href {https://doi.org/10.1038/nphoton.2017.40} {\bibfield
  {journal} {\bibinfo  {journal} {Nature Photonics}\ }\textbf {\bibinfo
  {volume} {11}},\ \bibinfo {pages} {309} (\bibinfo {year} {2017})}\BibitemShut
  {NoStop}%
\bibitem [{\citenamefont {Pompili}\ \emph {et~al.}(2021)\citenamefont
  {Pompili}, \citenamefont {Hermans}, \citenamefont {Baier}, \citenamefont
  {Beukers}, \citenamefont {Humphreys}, \citenamefont {Schouten}, \citenamefont
  {Vermeulen}, \citenamefont {Tiggelman}, \citenamefont {{dos Santos Martins}},
  \citenamefont {Dirkse}, \citenamefont {Wehner},\ and\ \citenamefont
  {Hanson}}]{Pompili2021}%
  \BibitemOpen
  \bibfield  {author} {\bibinfo {author} {\bibfnamefont {M.}~\bibnamefont
  {Pompili}}, \bibinfo {author} {\bibfnamefont {S.~L.~N.}\ \bibnamefont
  {Hermans}}, \bibinfo {author} {\bibfnamefont {S.}~\bibnamefont {Baier}},
  \bibinfo {author} {\bibfnamefont {H.~K.~C.}\ \bibnamefont {Beukers}},
  \bibinfo {author} {\bibfnamefont {P.~C.}\ \bibnamefont {Humphreys}}, \bibinfo
  {author} {\bibfnamefont {R.~N.}\ \bibnamefont {Schouten}}, \bibinfo {author}
  {\bibfnamefont {R.~F.~L.}\ \bibnamefont {Vermeulen}}, \bibinfo {author}
  {\bibfnamefont {M.~J.}\ \bibnamefont {Tiggelman}}, \bibinfo {author}
  {\bibfnamefont {L.}~\bibnamefont {{dos Santos Martins}}}, \bibinfo {author}
  {\bibfnamefont {B.}~\bibnamefont {Dirkse}}, \bibinfo {author} {\bibfnamefont
  {S.}~\bibnamefont {Wehner}},\ and\ \bibinfo {author} {\bibfnamefont
  {R.}~\bibnamefont {Hanson}},\ }\bibfield  {title} {\bibinfo {title}
  {{Realization of a multinode quantum network of remote solid-state qubits}},\
  }\href {https://doi.org/10.1126/science.abg1919} {\bibfield  {journal}
  {\bibinfo  {journal} {Science}\ }\textbf {\bibinfo {volume} {372}},\ \bibinfo
  {pages} {259} (\bibinfo {year} {2021})}\BibitemShut {NoStop}%
\bibitem [{\citenamefont {Bhaskar}\ \emph {et~al.}(2020)\citenamefont
  {Bhaskar}, \citenamefont {Riedinger}, \citenamefont {Machielse},
  \citenamefont {Levonian}, \citenamefont {Nguyen}, \citenamefont {Knall},
  \citenamefont {Park}, \citenamefont {Englund}, \citenamefont {Lon{\v{c}}ar},
  \citenamefont {Sukachev},\ and\ \citenamefont {Lukin}}]{Bhaskar2020}%
  \BibitemOpen
  \bibfield  {author} {\bibinfo {author} {\bibfnamefont {M.~K.}\ \bibnamefont
  {Bhaskar}}, \bibinfo {author} {\bibfnamefont {R.}~\bibnamefont {Riedinger}},
  \bibinfo {author} {\bibfnamefont {B.}~\bibnamefont {Machielse}}, \bibinfo
  {author} {\bibfnamefont {D.~S.}\ \bibnamefont {Levonian}}, \bibinfo {author}
  {\bibfnamefont {C.~T.}\ \bibnamefont {Nguyen}}, \bibinfo {author}
  {\bibfnamefont {E.~N.}\ \bibnamefont {Knall}}, \bibinfo {author}
  {\bibfnamefont {H.}~\bibnamefont {Park}}, \bibinfo {author} {\bibfnamefont
  {D.}~\bibnamefont {Englund}}, \bibinfo {author} {\bibfnamefont
  {M.}~\bibnamefont {Lon{\v{c}}ar}}, \bibinfo {author} {\bibfnamefont {D.~D.}\
  \bibnamefont {Sukachev}},\ and\ \bibinfo {author} {\bibfnamefont {M.~D.}\
  \bibnamefont {Lukin}},\ }\bibfield  {title} {\bibinfo {title} {{Experimental
  demonstration of memory-enhanced quantum communication}},\ }\href
  {https://doi.org/10.1038/s41586-020-2103-5} {\bibfield  {journal} {\bibinfo
  {journal} {Nature}\ }\textbf {\bibinfo {volume} {580}},\ \bibinfo {pages}
  {60} (\bibinfo {year} {2020})}\BibitemShut {NoStop}%
\bibitem [{\citenamefont {Knaut}\ \emph {et~al.}(2024)\citenamefont {Knaut},
  \citenamefont {Suleymanzade}, \citenamefont {Wei}, \citenamefont {Assumpcao},
  \citenamefont {Stas}, \citenamefont {Huan}, \citenamefont {Machielse},
  \citenamefont {Knall}, \citenamefont {Sutula}, \citenamefont {Baranes},
  \citenamefont {Sinclair}, \citenamefont {De-Eknamkul}, \citenamefont
  {Levonian}, \citenamefont {Bhaskar}, \citenamefont {Park}, \citenamefont
  {Lon{\v{c}}ar},\ and\ \citenamefont {Lukin}}]{Knaut2024}%
  \BibitemOpen
  \bibfield  {author} {\bibinfo {author} {\bibfnamefont {C.~M.}\ \bibnamefont
  {Knaut}}, \bibinfo {author} {\bibfnamefont {A.}~\bibnamefont {Suleymanzade}},
  \bibinfo {author} {\bibfnamefont {Y.~C.}\ \bibnamefont {Wei}}, \bibinfo
  {author} {\bibfnamefont {D.~R.}\ \bibnamefont {Assumpcao}}, \bibinfo {author}
  {\bibfnamefont {P.~J.}\ \bibnamefont {Stas}}, \bibinfo {author}
  {\bibfnamefont {Y.~Q.}\ \bibnamefont {Huan}}, \bibinfo {author}
  {\bibfnamefont {B.}~\bibnamefont {Machielse}}, \bibinfo {author}
  {\bibfnamefont {E.~N.}\ \bibnamefont {Knall}}, \bibinfo {author}
  {\bibfnamefont {M.}~\bibnamefont {Sutula}}, \bibinfo {author} {\bibfnamefont
  {G.}~\bibnamefont {Baranes}}, \bibinfo {author} {\bibfnamefont
  {N.}~\bibnamefont {Sinclair}}, \bibinfo {author} {\bibfnamefont
  {C.}~\bibnamefont {De-Eknamkul}}, \bibinfo {author} {\bibfnamefont {D.~S.}\
  \bibnamefont {Levonian}}, \bibinfo {author} {\bibfnamefont {M.~K.}\
  \bibnamefont {Bhaskar}}, \bibinfo {author} {\bibfnamefont {H.}~\bibnamefont
  {Park}}, \bibinfo {author} {\bibfnamefont {M.}~\bibnamefont {Lon{\v{c}}ar}},\
  and\ \bibinfo {author} {\bibfnamefont {M.~D.}\ \bibnamefont {Lukin}},\
  }\bibfield  {title} {\bibinfo {title} {{Entanglement of nanophotonic quantum
  memory nodes in a telecom network}},\ }\href
  {https://doi.org/10.1038/s41586-024-07252-z} {\bibfield  {journal} {\bibinfo
  {journal} {Nature}\ }\textbf {\bibinfo {volume} {629}},\ \bibinfo {pages}
  {573} (\bibinfo {year} {2024})}\BibitemShut {NoStop}%
\bibitem [{\citenamefont {Stolk}\ \emph {et~al.}(2024)\citenamefont {Stolk},
  \citenamefont {van~der Enden}, \citenamefont {Slater}, \citenamefont
  {te~Raa-Derckx}, \citenamefont {Botma}, \citenamefont {van Rantwijk},
  \citenamefont {Biemond}, \citenamefont {Hagen}, \citenamefont {Herfst},
  \citenamefont {Koek}, \citenamefont {Meskers}, \citenamefont {Vollmer},
  \citenamefont {van Zwet}, \citenamefont {Markham}, \citenamefont {Edmonds},
  \citenamefont {Geus}, \citenamefont {Elsen}, \citenamefont {Jungbluth},
  \citenamefont {Haefner}, \citenamefont {Tresp}, \citenamefont {Stuhler},
  \citenamefont {Ritter},\ and\ \citenamefont {Hanson}}]{Stolk2024}%
  \BibitemOpen
  \bibfield  {author} {\bibinfo {author} {\bibfnamefont {A.~J.}\ \bibnamefont
  {Stolk}}, \bibinfo {author} {\bibfnamefont {K.~L.}\ \bibnamefont {van~der
  Enden}}, \bibinfo {author} {\bibfnamefont {M.-C.}\ \bibnamefont {Slater}},
  \bibinfo {author} {\bibfnamefont {I.}~\bibnamefont {te~Raa-Derckx}}, \bibinfo
  {author} {\bibfnamefont {P.}~\bibnamefont {Botma}}, \bibinfo {author}
  {\bibfnamefont {J.}~\bibnamefont {van Rantwijk}}, \bibinfo {author}
  {\bibfnamefont {J.~J.~B.}\ \bibnamefont {Biemond}}, \bibinfo {author}
  {\bibfnamefont {R.~A.~J.}\ \bibnamefont {Hagen}}, \bibinfo {author}
  {\bibfnamefont {R.~W.}\ \bibnamefont {Herfst}}, \bibinfo {author}
  {\bibfnamefont {W.~D.}\ \bibnamefont {Koek}}, \bibinfo {author}
  {\bibfnamefont {A.~J.~H.}\ \bibnamefont {Meskers}}, \bibinfo {author}
  {\bibfnamefont {R.}~\bibnamefont {Vollmer}}, \bibinfo {author} {\bibfnamefont
  {E.~J.}\ \bibnamefont {van Zwet}}, \bibinfo {author} {\bibfnamefont
  {M.}~\bibnamefont {Markham}}, \bibinfo {author} {\bibfnamefont {A.~M.}\
  \bibnamefont {Edmonds}}, \bibinfo {author} {\bibfnamefont {J.~F.}\
  \bibnamefont {Geus}}, \bibinfo {author} {\bibfnamefont {F.}~\bibnamefont
  {Elsen}}, \bibinfo {author} {\bibfnamefont {B.}~\bibnamefont {Jungbluth}},
  \bibinfo {author} {\bibfnamefont {C.}~\bibnamefont {Haefner}}, \bibinfo
  {author} {\bibfnamefont {C.}~\bibnamefont {Tresp}}, \bibinfo {author}
  {\bibfnamefont {J.}~\bibnamefont {Stuhler}}, \bibinfo {author} {\bibfnamefont
  {S.}~\bibnamefont {Ritter}},\ and\ \bibinfo {author} {\bibfnamefont
  {R.}~\bibnamefont {Hanson}},\ }\bibfield  {title} {\bibinfo {title}
  {{Metropolitan-scale heralded entanglement of solid-state qubits}},\ }\href
  {http://arxiv.org/abs/2404.03723} {\bibfield  {journal} {\bibinfo  {journal}
  {arXiv:2404.03723}\ } (\bibinfo {year} {2024})}\BibitemShut {NoStop}%
\bibitem [{\citenamefont {Manson}\ and\ \citenamefont
  {Harrison}(2005)}]{Manson2005}%
  \BibitemOpen
  \bibfield  {author} {\bibinfo {author} {\bibfnamefont {N.~B.}\ \bibnamefont
  {Manson}}\ and\ \bibinfo {author} {\bibfnamefont {J.~P.}\ \bibnamefont
  {Harrison}},\ }\bibfield  {title} {\bibinfo {title} {{Photo-ionization of the
  nitrogen-vacancy center in diamond}},\ }\href
  {https://doi.org/10.1016/j.diamond.2005.06.027} {\bibfield  {journal}
  {\bibinfo  {journal} {Diamond and Related Materials}\ }\textbf {\bibinfo
  {volume} {14}},\ \bibinfo {pages} {1705} (\bibinfo {year}
  {2005})}\BibitemShut {NoStop}%
\bibitem [{\citenamefont {Jelezko}\ and\ \citenamefont
  {Wrachtrup}(2006)}]{Jelezko2006}%
  \BibitemOpen
  \bibfield  {author} {\bibinfo {author} {\bibfnamefont {F.}~\bibnamefont
  {Jelezko}}\ and\ \bibinfo {author} {\bibfnamefont {J.}~\bibnamefont
  {Wrachtrup}},\ }\bibfield  {title} {\bibinfo {title} {{Single defect centres
  in diamond: A review}},\ }\href {https://doi.org/10.1002/pssa.200671403}
  {\bibfield  {journal} {\bibinfo  {journal} {Physica Status Solidi (A)}\
  }\textbf {\bibinfo {volume} {203}},\ \bibinfo {pages} {3207} (\bibinfo {year}
  {2006})}\BibitemShut {NoStop}%
\bibitem [{\citenamefont {Childress}\ \emph {et~al.}(2006)\citenamefont
  {Childress}, \citenamefont {{Gurudev Dutt}}, \citenamefont {Taylor},
  \citenamefont {Zibrov}, \citenamefont {Jelezko}, \citenamefont {Wrachtrup},
  \citenamefont {Hemmer},\ and\ \citenamefont {Lukin}}]{Childress2006}%
  \BibitemOpen
  \bibfield  {author} {\bibinfo {author} {\bibfnamefont {L.}~\bibnamefont
  {Childress}}, \bibinfo {author} {\bibfnamefont {M.~V.}\ \bibnamefont
  {{Gurudev Dutt}}}, \bibinfo {author} {\bibfnamefont {J.~M.}\ \bibnamefont
  {Taylor}}, \bibinfo {author} {\bibfnamefont {A.~S.}\ \bibnamefont {Zibrov}},
  \bibinfo {author} {\bibfnamefont {F.}~\bibnamefont {Jelezko}}, \bibinfo
  {author} {\bibfnamefont {J.}~\bibnamefont {Wrachtrup}}, \bibinfo {author}
  {\bibfnamefont {P.~R.}\ \bibnamefont {Hemmer}},\ and\ \bibinfo {author}
  {\bibfnamefont {M.~D.}\ \bibnamefont {Lukin}},\ }\bibfield  {title} {\bibinfo
  {title} {{Coherent dynamics of coupled electron and nuclear spin qubits in
  diamond}},\ }\href {https://doi.org/10.1126/science.1131871} {\bibfield
  {journal} {\bibinfo  {journal} {Science}\ }\textbf {\bibinfo {volume}
  {314}},\ \bibinfo {pages} {281} (\bibinfo {year} {2006})}\BibitemShut
  {NoStop}%
\bibitem [{\citenamefont {Dolde}\ \emph {et~al.}(2011)\citenamefont {Dolde},
  \citenamefont {Fedder}, \citenamefont {Doherty}, \citenamefont
  {N{\"{o}}bauer}, \citenamefont {Rempp}, \citenamefont {Balasubramanian},
  \citenamefont {Wolf}, \citenamefont {Reinhard}, \citenamefont {Hollenberg},
  \citenamefont {Jelezko},\ and\ \citenamefont {Wrachtrup}}]{Dolde2011}%
  \BibitemOpen
  \bibfield  {author} {\bibinfo {author} {\bibfnamefont {F.}~\bibnamefont
  {Dolde}}, \bibinfo {author} {\bibfnamefont {H.}~\bibnamefont {Fedder}},
  \bibinfo {author} {\bibfnamefont {M.~W.}\ \bibnamefont {Doherty}}, \bibinfo
  {author} {\bibfnamefont {T.}~\bibnamefont {N{\"{o}}bauer}}, \bibinfo {author}
  {\bibfnamefont {F.}~\bibnamefont {Rempp}}, \bibinfo {author} {\bibfnamefont
  {G.}~\bibnamefont {Balasubramanian}}, \bibinfo {author} {\bibfnamefont
  {T.}~\bibnamefont {Wolf}}, \bibinfo {author} {\bibfnamefont {F.}~\bibnamefont
  {Reinhard}}, \bibinfo {author} {\bibfnamefont {L.~C.~L.}\ \bibnamefont
  {Hollenberg}}, \bibinfo {author} {\bibfnamefont {F.}~\bibnamefont
  {Jelezko}},\ and\ \bibinfo {author} {\bibfnamefont {J.}~\bibnamefont
  {Wrachtrup}},\ }\bibfield  {title} {\bibinfo {title} {{Electric-field sensing
  using single diamond spins}},\ }\href {https://doi.org/10.1038/nphys1969}
  {\bibfield  {journal} {\bibinfo  {journal} {Nature Physics}\ }\textbf
  {\bibinfo {volume} {7}},\ \bibinfo {pages} {459} (\bibinfo {year}
  {2011})}\BibitemShut {NoStop}%
\bibitem [{\citenamefont {Ruf}\ \emph {et~al.}(2021)\citenamefont {Ruf},
  \citenamefont {Wan}, \citenamefont {Choi}, \citenamefont {Englund},\ and\
  \citenamefont {Hanson}}]{Ruf2021}%
  \BibitemOpen
  \bibfield  {author} {\bibinfo {author} {\bibfnamefont {M.}~\bibnamefont
  {Ruf}}, \bibinfo {author} {\bibfnamefont {N.~H.}\ \bibnamefont {Wan}},
  \bibinfo {author} {\bibfnamefont {H.}~\bibnamefont {Choi}}, \bibinfo {author}
  {\bibfnamefont {D.}~\bibnamefont {Englund}},\ and\ \bibinfo {author}
  {\bibfnamefont {R.}~\bibnamefont {Hanson}},\ }\bibfield  {title} {\bibinfo
  {title} {{Quantum networks based on color centers in diamond}},\ }\href
  {https://doi.org/10.1063/5.0056534} {\bibfield  {journal} {\bibinfo
  {journal} {Journal of Applied Physics}\ }\textbf {\bibinfo {volume} {130}},\
  \bibinfo {pages} {070901} (\bibinfo {year} {2021})}\BibitemShut {NoStop}%
\bibitem [{\citenamefont {M{\"{u}}ller}\ \emph {et~al.}(2014)\citenamefont
  {M{\"{u}}ller}, \citenamefont {Hepp}, \citenamefont {Pingault}, \citenamefont
  {Neu}, \citenamefont {Gsell}, \citenamefont {Schreck}, \citenamefont
  {Sternschulte}, \citenamefont {Steinm{\"{u}}ller-Nethl}, \citenamefont
  {Becher},\ and\ \citenamefont {Atat{\"{u}}re}}]{Mu2014}%
  \BibitemOpen
  \bibfield  {author} {\bibinfo {author} {\bibfnamefont {T.}~\bibnamefont
  {M{\"{u}}ller}}, \bibinfo {author} {\bibfnamefont {C.}~\bibnamefont {Hepp}},
  \bibinfo {author} {\bibfnamefont {B.}~\bibnamefont {Pingault}}, \bibinfo
  {author} {\bibfnamefont {E.}~\bibnamefont {Neu}}, \bibinfo {author}
  {\bibfnamefont {S.}~\bibnamefont {Gsell}}, \bibinfo {author} {\bibfnamefont
  {M.}~\bibnamefont {Schreck}}, \bibinfo {author} {\bibfnamefont
  {H.}~\bibnamefont {Sternschulte}}, \bibinfo {author} {\bibfnamefont
  {D.}~\bibnamefont {Steinm{\"{u}}ller-Nethl}}, \bibinfo {author}
  {\bibfnamefont {C.}~\bibnamefont {Becher}},\ and\ \bibinfo {author}
  {\bibfnamefont {M.}~\bibnamefont {Atat{\"{u}}re}},\ }\bibfield  {title}
  {\bibinfo {title} {{Optical signatures of silicon-vacancy spins in
  diamond}},\ }\href {https://doi.org/10.1038/ncomms4328} {\bibfield  {journal}
  {\bibinfo  {journal} {Nature Communications}\ }\textbf {\bibinfo {volume}
  {5}},\ \bibinfo {pages} {3328} (\bibinfo {year} {2014})}\BibitemShut
  {NoStop}%
\bibitem [{\citenamefont {Rogers}\ \emph {et~al.}(2014)\citenamefont {Rogers},
  \citenamefont {Jahnke}, \citenamefont {Metsch}, \citenamefont {Sipahigil},
  \citenamefont {Binder}, \citenamefont {Teraji}, \citenamefont {Sumiya},
  \citenamefont {Isoya}, \citenamefont {Lukin}, \citenamefont {Hemmer},\ and\
  \citenamefont {Jelezko}}]{Rogers2014a}%
  \BibitemOpen
  \bibfield  {author} {\bibinfo {author} {\bibfnamefont {L.~J.}\ \bibnamefont
  {Rogers}}, \bibinfo {author} {\bibfnamefont {K.~D.}\ \bibnamefont {Jahnke}},
  \bibinfo {author} {\bibfnamefont {M.~H.}\ \bibnamefont {Metsch}}, \bibinfo
  {author} {\bibfnamefont {A.}~\bibnamefont {Sipahigil}}, \bibinfo {author}
  {\bibfnamefont {J.~M.}\ \bibnamefont {Binder}}, \bibinfo {author}
  {\bibfnamefont {T.}~\bibnamefont {Teraji}}, \bibinfo {author} {\bibfnamefont
  {H.}~\bibnamefont {Sumiya}}, \bibinfo {author} {\bibfnamefont
  {J.}~\bibnamefont {Isoya}}, \bibinfo {author} {\bibfnamefont {M.~D.}\
  \bibnamefont {Lukin}}, \bibinfo {author} {\bibfnamefont {P.}~\bibnamefont
  {Hemmer}},\ and\ \bibinfo {author} {\bibfnamefont {F.}~\bibnamefont
  {Jelezko}},\ }\bibfield  {title} {\bibinfo {title} {{All-Optical
  Initialization, Readout, and Coherent Preparation of Single Silicon-Vacancy
  Spins in Diamond}},\ }\href {https://doi.org/10.1103/PhysRevLett.113.263602}
  {\bibfield  {journal} {\bibinfo  {journal} {Physical Review Letters}\
  }\textbf {\bibinfo {volume} {113}},\ \bibinfo {pages} {263602} (\bibinfo
  {year} {2014})}\BibitemShut {NoStop}%
\bibitem [{\citenamefont {Becker}\ \emph {et~al.}(2018)\citenamefont {Becker},
  \citenamefont {Pingault}, \citenamefont {Gro{\ss}}, \citenamefont
  {G{\"{u}}ndo^^c4^^9fan}, \citenamefont {Kukharchyk}, \citenamefont {Markham},
  \citenamefont {Edmonds}, \citenamefont {Atat{\"{u}}re}, \citenamefont
  {Bushev},\ and\ \citenamefont {Becher}}]{Becker2018a}%
  \BibitemOpen
  \bibfield  {author} {\bibinfo {author} {\bibfnamefont {J.~N.}\ \bibnamefont
  {Becker}}, \bibinfo {author} {\bibfnamefont {B.}~\bibnamefont {Pingault}},
  \bibinfo {author} {\bibfnamefont {D.}~\bibnamefont {Gro{\ss}}}, \bibinfo
  {author} {\bibfnamefont {M.}~\bibnamefont {G{\"{u}}ndo^^c4^^9fan}}, \bibinfo
  {author} {\bibfnamefont {N.}~\bibnamefont {Kukharchyk}}, \bibinfo {author}
  {\bibfnamefont {M.}~\bibnamefont {Markham}}, \bibinfo {author} {\bibfnamefont
  {A.}~\bibnamefont {Edmonds}}, \bibinfo {author} {\bibfnamefont
  {M.}~\bibnamefont {Atat{\"{u}}re}}, \bibinfo {author} {\bibfnamefont
  {P.}~\bibnamefont {Bushev}},\ and\ \bibinfo {author} {\bibfnamefont
  {C.}~\bibnamefont {Becher}},\ }\bibfield  {title} {\bibinfo {title}
  {{All-Optical Control of the Silicon-Vacancy Spin in Diamond at Millikelvin
  Temperatures}},\ }\href {https://doi.org/10.1103/PhysRevLett.120.053603}
  {\bibfield  {journal} {\bibinfo  {journal} {Physical Review Letters}\
  }\textbf {\bibinfo {volume} {120}},\ \bibinfo {pages} {053603} (\bibinfo
  {year} {2018})}\BibitemShut {NoStop}%
\bibitem [{\citenamefont {Iwasaki}\ \emph {et~al.}(2017)\citenamefont
  {Iwasaki}, \citenamefont {Miyamoto}, \citenamefont {Taniguchi}, \citenamefont
  {Siyushev}, \citenamefont {Metsch}, \citenamefont {Jelezko},\ and\
  \citenamefont {Hatano}}]{Iwasaki2017}%
  \BibitemOpen
  \bibfield  {author} {\bibinfo {author} {\bibfnamefont {T.}~\bibnamefont
  {Iwasaki}}, \bibinfo {author} {\bibfnamefont {Y.}~\bibnamefont {Miyamoto}},
  \bibinfo {author} {\bibfnamefont {T.}~\bibnamefont {Taniguchi}}, \bibinfo
  {author} {\bibfnamefont {P.}~\bibnamefont {Siyushev}}, \bibinfo {author}
  {\bibfnamefont {M.~H.}\ \bibnamefont {Metsch}}, \bibinfo {author}
  {\bibfnamefont {F.}~\bibnamefont {Jelezko}},\ and\ \bibinfo {author}
  {\bibfnamefont {M.}~\bibnamefont {Hatano}},\ }\bibfield  {title} {\bibinfo
  {title} {{Tin-Vacancy Quantum Emitters in Diamond}},\ }\href
  {https://doi.org/10.1103/PhysRevLett.119.253601} {\bibfield  {journal}
  {\bibinfo  {journal} {Physical Review Letters}\ }\textbf {\bibinfo {volume}
  {119}},\ \bibinfo {pages} {253601} (\bibinfo {year} {2017})}\BibitemShut
  {NoStop}%
\bibitem [{\citenamefont {H{\"{a}}u{\ss}ler}\ \emph {et~al.}(2017)\citenamefont
  {H{\"{a}}u{\ss}ler}, \citenamefont {Thiering}, \citenamefont {Dietrich},
  \citenamefont {Waasem}, \citenamefont {Teraji}, \citenamefont {Isoya},
  \citenamefont {Iwasaki}, \citenamefont {Hatano}, \citenamefont {Jelezko},
  \citenamefont {Gali},\ and\ \citenamefont {Kubanek}}]{Hau^^c3^^9fler2017}%
  \BibitemOpen
  \bibfield  {author} {\bibinfo {author} {\bibfnamefont {S.}~\bibnamefont
  {H{\"{a}}u{\ss}ler}}, \bibinfo {author} {\bibfnamefont {G.}~\bibnamefont
  {Thiering}}, \bibinfo {author} {\bibfnamefont {A.}~\bibnamefont {Dietrich}},
  \bibinfo {author} {\bibfnamefont {N.}~\bibnamefont {Waasem}}, \bibinfo
  {author} {\bibfnamefont {T.}~\bibnamefont {Teraji}}, \bibinfo {author}
  {\bibfnamefont {J.}~\bibnamefont {Isoya}}, \bibinfo {author} {\bibfnamefont
  {T.}~\bibnamefont {Iwasaki}}, \bibinfo {author} {\bibfnamefont
  {M.}~\bibnamefont {Hatano}}, \bibinfo {author} {\bibfnamefont
  {F.}~\bibnamefont {Jelezko}}, \bibinfo {author} {\bibfnamefont
  {A.}~\bibnamefont {Gali}},\ and\ \bibinfo {author} {\bibfnamefont
  {A.}~\bibnamefont {Kubanek}},\ }\bibfield  {title} {\bibinfo {title}
  {{Photoluminescence excitation spectroscopy of SiV$^-$ and GeV$^-$ color
  center in diamond}},\ }\href {https://doi.org/10.1088/1367-2630/aa73e5}
  {\bibfield  {journal} {\bibinfo  {journal} {New Journal of Physics}\ }\textbf
  {\bibinfo {volume} {19}},\ \bibinfo {pages} {063036} (\bibinfo {year}
  {2017})}\BibitemShut {NoStop}%
\bibitem [{\citenamefont {Bradac}\ \emph {et~al.}(2019)\citenamefont {Bradac},
  \citenamefont {Gao}, \citenamefont {Forneris}, \citenamefont {Trusheim},\
  and\ \citenamefont {Aharonovich}}]{Bradac2019}%
  \BibitemOpen
  \bibfield  {author} {\bibinfo {author} {\bibfnamefont {C.}~\bibnamefont
  {Bradac}}, \bibinfo {author} {\bibfnamefont {W.}~\bibnamefont {Gao}},
  \bibinfo {author} {\bibfnamefont {J.}~\bibnamefont {Forneris}}, \bibinfo
  {author} {\bibfnamefont {M.~E.}\ \bibnamefont {Trusheim}},\ and\ \bibinfo
  {author} {\bibfnamefont {I.}~\bibnamefont {Aharonovich}},\ }\bibfield
  {title} {\bibinfo {title} {{Quantum nanophotonics with group IV defects in
  diamond}},\ }\href {https://doi.org/10.1038/s41467-019-13332-w} {\bibfield
  {journal} {\bibinfo  {journal} {Nature Communications}\ }\textbf {\bibinfo
  {volume} {10}},\ \bibinfo {pages} {5625} (\bibinfo {year}
  {2019})}\BibitemShut {NoStop}%
\bibitem [{\citenamefont {Chen}\ \emph {et~al.}(2020)\citenamefont {Chen},
  \citenamefont {Zheludev},\ and\ \citenamefont {Gao}}]{Chen2020b}%
  \BibitemOpen
  \bibfield  {author} {\bibinfo {author} {\bibfnamefont {D.}~\bibnamefont
  {Chen}}, \bibinfo {author} {\bibfnamefont {N.}~\bibnamefont {Zheludev}},\
  and\ \bibinfo {author} {\bibfnamefont {W.}~\bibnamefont {Gao}},\ }\bibfield
  {title} {\bibinfo {title} {{Building Blocks for Quantum Network Based on
  Group-IV Split-Vacancy Centers in Diamond}},\ }\href
  {https://doi.org/10.1002/qute.201900069} {\bibfield  {journal} {\bibinfo
  {journal} {Advanced Quantum Technologies}\ }\textbf {\bibinfo {volume} {3}},\
  \bibinfo {pages} {1900069} (\bibinfo {year} {2020})}\BibitemShut {NoStop}%
\bibitem [{\citenamefont {Parker}\ \emph {et~al.}(2024)\citenamefont {Parker},
  \citenamefont {{Arjona Mart{\'{i}}nez}}, \citenamefont {Chen}, \citenamefont
  {Stramma}, \citenamefont {Harris}, \citenamefont {Michaels}, \citenamefont
  {Trusheim}, \citenamefont {{Hayhurst Appel}}, \citenamefont {Purser},
  \citenamefont {Roth}, \citenamefont {Englund},\ and\ \citenamefont
  {Atat{\"{u}}re}}]{Parker2024}%
  \BibitemOpen
  \bibfield  {author} {\bibinfo {author} {\bibfnamefont {R.~A.}\ \bibnamefont
  {Parker}}, \bibinfo {author} {\bibfnamefont {J.}~\bibnamefont {{Arjona
  Mart{\'{i}}nez}}}, \bibinfo {author} {\bibfnamefont {K.~C.}\ \bibnamefont
  {Chen}}, \bibinfo {author} {\bibfnamefont {A.~M.}\ \bibnamefont {Stramma}},
  \bibinfo {author} {\bibfnamefont {I.~B.}\ \bibnamefont {Harris}}, \bibinfo
  {author} {\bibfnamefont {C.~P.}\ \bibnamefont {Michaels}}, \bibinfo {author}
  {\bibfnamefont {M.~E.}\ \bibnamefont {Trusheim}}, \bibinfo {author}
  {\bibfnamefont {M.}~\bibnamefont {{Hayhurst Appel}}}, \bibinfo {author}
  {\bibfnamefont {C.~M.}\ \bibnamefont {Purser}}, \bibinfo {author}
  {\bibfnamefont {W.~G.}\ \bibnamefont {Roth}}, \bibinfo {author}
  {\bibfnamefont {D.}~\bibnamefont {Englund}},\ and\ \bibinfo {author}
  {\bibfnamefont {M.}~\bibnamefont {Atat{\"{u}}re}},\ }\bibfield  {title}
  {\bibinfo {title} {{A diamond nanophotonic interface with an optically
  accessible deterministic electronuclear spin register}},\ }\href
  {https://doi.org/10.1038/s41566-023-01332-8} {\bibfield  {journal} {\bibinfo
  {journal} {Nature Photonics}\ }\textbf {\bibinfo {volume} {18}},\ \bibinfo
  {pages} {156} (\bibinfo {year} {2024})}\BibitemShut {NoStop}%
\bibitem [{\citenamefont {Senkalla}\ \emph {et~al.}(2024)\citenamefont
  {Senkalla}, \citenamefont {Genov}, \citenamefont {Metsch}, \citenamefont
  {Siyushev},\ and\ \citenamefont {Jelezko}}]{Senkalla2024}%
  \BibitemOpen
  \bibfield  {author} {\bibinfo {author} {\bibfnamefont {K.}~\bibnamefont
  {Senkalla}}, \bibinfo {author} {\bibfnamefont {G.}~\bibnamefont {Genov}},
  \bibinfo {author} {\bibfnamefont {M.~H.}\ \bibnamefont {Metsch}}, \bibinfo
  {author} {\bibfnamefont {P.}~\bibnamefont {Siyushev}},\ and\ \bibinfo
  {author} {\bibfnamefont {F.}~\bibnamefont {Jelezko}},\ }\bibfield  {title}
  {\bibinfo {title} {{Germanium Vacancy in Diamond Quantum Memory Exceeding 20
  ms}},\ }\href {https://doi.org/10.1103/PhysRevLett.132.026901} {\bibfield
  {journal} {\bibinfo  {journal} {Physical Review Letters}\ }\textbf {\bibinfo
  {volume} {132}},\ \bibinfo {pages} {026901} (\bibinfo {year}
  {2024})}\BibitemShut {NoStop}%
\bibitem [{\citenamefont {Kurokawa}\ \emph {et~al.}(2024)\citenamefont
  {Kurokawa}, \citenamefont {Wakamatsu}, \citenamefont {Nakazato},
  \citenamefont {Makino}, \citenamefont {Kato}, \citenamefont {Sekiguchi},\
  and\ \citenamefont {Kosaka}}]{Kurokawa2024}%
  \BibitemOpen
  \bibfield  {author} {\bibinfo {author} {\bibfnamefont {H.}~\bibnamefont
  {Kurokawa}}, \bibinfo {author} {\bibfnamefont {K.}~\bibnamefont {Wakamatsu}},
  \bibinfo {author} {\bibfnamefont {S.}~\bibnamefont {Nakazato}}, \bibinfo
  {author} {\bibfnamefont {T.}~\bibnamefont {Makino}}, \bibinfo {author}
  {\bibfnamefont {H.}~\bibnamefont {Kato}}, \bibinfo {author} {\bibfnamefont
  {Y.}~\bibnamefont {Sekiguchi}},\ and\ \bibinfo {author} {\bibfnamefont
  {H.}~\bibnamefont {Kosaka}},\ }\bibfield  {title} {\bibinfo {title}
  {{Coherent electric field control of orbital state of a neutral
  nitrogen-vacancy center}},\ }\href
  {https://doi.org/10.1038/s41467-024-47973-3} {\bibfield  {journal} {\bibinfo
  {journal} {Nature Communications}\ }\textbf {\bibinfo {volume} {15}},\
  \bibinfo {pages} {4039} (\bibinfo {year} {2024})}\BibitemShut {NoStop}%
\bibitem [{\citenamefont {Venturi}\ \emph {et~al.}(2019)\citenamefont
  {Venturi}, \citenamefont {Rigutti}, \citenamefont {Houard}, \citenamefont
  {Blum}, \citenamefont {Malykhin}, \citenamefont {Obraztsov},\ and\
  \citenamefont {Vella}}]{Venturi2019}%
  \BibitemOpen
  \bibfield  {author} {\bibinfo {author} {\bibfnamefont {L.}~\bibnamefont
  {Venturi}}, \bibinfo {author} {\bibfnamefont {L.}~\bibnamefont {Rigutti}},
  \bibinfo {author} {\bibfnamefont {J.}~\bibnamefont {Houard}}, \bibinfo
  {author} {\bibfnamefont {I.}~\bibnamefont {Blum}}, \bibinfo {author}
  {\bibfnamefont {S.}~\bibnamefont {Malykhin}}, \bibinfo {author}
  {\bibfnamefont {A.}~\bibnamefont {Obraztsov}},\ and\ \bibinfo {author}
  {\bibfnamefont {A.}~\bibnamefont {Vella}},\ }\bibfield  {title} {\bibinfo
  {title} {{Strain sensitivity and symmetry of 2.65 eV color center in diamond
  nanoscale needles}},\ }\href {https://doi.org/10.1063/1.5092329} {\bibfield
  {journal} {\bibinfo  {journal} {Applied Physics Letters}\ }\textbf {\bibinfo
  {volume} {114}},\ \bibinfo {pages} {143104} (\bibinfo {year}
  {2019})}\BibitemShut {NoStop}%
\bibitem [{\citenamefont {{Van Oort}}\ and\ \citenamefont
  {Glasbeek}(1990)}]{Oort1990}%
  \BibitemOpen
  \bibfield  {author} {\bibinfo {author} {\bibfnamefont {E.}~\bibnamefont {{Van
  Oort}}}\ and\ \bibinfo {author} {\bibfnamefont {M.}~\bibnamefont
  {Glasbeek}},\ }\bibfield  {title} {\bibinfo {title} {{Electric-field-induced
  modulation of spin echoes of N-V centers in diamond}},\ }\href
  {https://doi.org/10.1016/0009-2614(90)85665-Y} {\bibfield  {journal}
  {\bibinfo  {journal} {Chemical Physics Letters}\ }\textbf {\bibinfo {volume}
  {168}},\ \bibinfo {pages} {529} (\bibinfo {year} {1990})}\BibitemShut
  {NoStop}%
\bibitem [{\citenamefont {Udvarhelyi}\ \emph {et~al.}(2018)\citenamefont
  {Udvarhelyi}, \citenamefont {Shkolnikov}, \citenamefont {Gali}, \citenamefont
  {Burkard},\ and\ \citenamefont {P{\'{a}}lyi}}]{Udvarhelyi2018}%
  \BibitemOpen
  \bibfield  {author} {\bibinfo {author} {\bibfnamefont {P.}~\bibnamefont
  {Udvarhelyi}}, \bibinfo {author} {\bibfnamefont {V.~O.}\ \bibnamefont
  {Shkolnikov}}, \bibinfo {author} {\bibfnamefont {A.}~\bibnamefont {Gali}},
  \bibinfo {author} {\bibfnamefont {G.}~\bibnamefont {Burkard}},\ and\ \bibinfo
  {author} {\bibfnamefont {A.}~\bibnamefont {P{\'{a}}lyi}},\ }\bibfield
  {title} {\bibinfo {title} {{Spin-strain interaction in nitrogen-vacancy
  centers in diamond}},\ }\href {https://doi.org/10.1103/PhysRevB.98.075201}
  {\bibfield  {journal} {\bibinfo  {journal} {Physical Review B}\ }\textbf
  {\bibinfo {volume} {98}},\ \bibinfo {pages} {075201} (\bibinfo {year}
  {2018})}\BibitemShut {NoStop}%
\bibitem [{\citenamefont {Baier}\ \emph {et~al.}(2020)\citenamefont {Baier},
  \citenamefont {Bradley}, \citenamefont {Middelburg}, \citenamefont
  {Dobrovitski}, \citenamefont {Taminiau},\ and\ \citenamefont
  {Hanson}}]{Baier2020}%
  \BibitemOpen
  \bibfield  {author} {\bibinfo {author} {\bibfnamefont {S.}~\bibnamefont
  {Baier}}, \bibinfo {author} {\bibfnamefont {C.~E.}\ \bibnamefont {Bradley}},
  \bibinfo {author} {\bibfnamefont {T.}~\bibnamefont {Middelburg}}, \bibinfo
  {author} {\bibfnamefont {V.~V.}\ \bibnamefont {Dobrovitski}}, \bibinfo
  {author} {\bibfnamefont {T.~H.}\ \bibnamefont {Taminiau}},\ and\ \bibinfo
  {author} {\bibfnamefont {R.}~\bibnamefont {Hanson}},\ }\bibfield  {title}
  {\bibinfo {title} {{Orbital and Spin Dynamics of Single Neutrally-Charged
  Nitrogen-Vacancy Centers in Diamond}},\ }\href
  {https://doi.org/10.1103/PhysRevLett.125.193601} {\bibfield  {journal}
  {\bibinfo  {journal} {Physical Review Letters}\ }\textbf {\bibinfo {volume}
  {125}},\ \bibinfo {pages} {193601} (\bibinfo {year} {2020})}\BibitemShut
  {NoStop}%
\bibitem [{\citenamefont {Nguyen}\ \emph {et~al.}(2019)\citenamefont {Nguyen},
  \citenamefont {Sukachev}, \citenamefont {Bhaskar}, \citenamefont {Machielse},
  \citenamefont {Levonian}, \citenamefont {Knall}, \citenamefont {Stroganov},
  \citenamefont {Riedinger}, \citenamefont {Park}, \citenamefont
  {Lon{\v{c}}ar},\ and\ \citenamefont {Lukin}}]{Nguyen2019}%
  \BibitemOpen
  \bibfield  {author} {\bibinfo {author} {\bibfnamefont {C.~T.}\ \bibnamefont
  {Nguyen}}, \bibinfo {author} {\bibfnamefont {D.~D.}\ \bibnamefont
  {Sukachev}}, \bibinfo {author} {\bibfnamefont {M.~K.}\ \bibnamefont
  {Bhaskar}}, \bibinfo {author} {\bibfnamefont {B.}~\bibnamefont {Machielse}},
  \bibinfo {author} {\bibfnamefont {D.~S.}\ \bibnamefont {Levonian}}, \bibinfo
  {author} {\bibfnamefont {E.~N.}\ \bibnamefont {Knall}}, \bibinfo {author}
  {\bibfnamefont {P.}~\bibnamefont {Stroganov}}, \bibinfo {author}
  {\bibfnamefont {R.}~\bibnamefont {Riedinger}}, \bibinfo {author}
  {\bibfnamefont {H.}~\bibnamefont {Park}}, \bibinfo {author} {\bibfnamefont
  {M.}~\bibnamefont {Lon{\v{c}}ar}},\ and\ \bibinfo {author} {\bibfnamefont
  {M.~D.}\ \bibnamefont {Lukin}},\ }\bibfield  {title} {\bibinfo {title}
  {{Quantum Network Nodes Based on Diamond Qubits with an Efficient
  Nanophotonic Interface}},\ }\href
  {https://doi.org/10.1103/PhysRevLett.123.183602} {\bibfield  {journal}
  {\bibinfo  {journal} {Physical Review Letters}\ }\textbf {\bibinfo {volume}
  {123}},\ \bibinfo {pages} {183602} (\bibinfo {year} {2019})}\BibitemShut
  {NoStop}%
\bibitem [{\citenamefont {Zhu}\ \emph {et~al.}(2011)\citenamefont {Zhu},
  \citenamefont {Saito}, \citenamefont {Kemp}, \citenamefont {Kakuyanagi},
  \citenamefont {Karimoto}, \citenamefont {Nakano}, \citenamefont {Munro},
  \citenamefont {Tokura}, \citenamefont {Everitt}, \citenamefont {Nemoto},
  \citenamefont {Kasu}, \citenamefont {Mizuochi},\ and\ \citenamefont
  {Semba}}]{Zhu2011}%
  \BibitemOpen
  \bibfield  {author} {\bibinfo {author} {\bibfnamefont {X.}~\bibnamefont
  {Zhu}}, \bibinfo {author} {\bibfnamefont {S.}~\bibnamefont {Saito}}, \bibinfo
  {author} {\bibfnamefont {A.}~\bibnamefont {Kemp}}, \bibinfo {author}
  {\bibfnamefont {K.}~\bibnamefont {Kakuyanagi}}, \bibinfo {author}
  {\bibfnamefont {S.}~\bibnamefont {Karimoto}}, \bibinfo {author}
  {\bibfnamefont {H.}~\bibnamefont {Nakano}}, \bibinfo {author} {\bibfnamefont
  {W.~J.}\ \bibnamefont {Munro}}, \bibinfo {author} {\bibfnamefont
  {Y.}~\bibnamefont {Tokura}}, \bibinfo {author} {\bibfnamefont {M.~S.}\
  \bibnamefont {Everitt}}, \bibinfo {author} {\bibfnamefont {K.}~\bibnamefont
  {Nemoto}}, \bibinfo {author} {\bibfnamefont {M.}~\bibnamefont {Kasu}},
  \bibinfo {author} {\bibfnamefont {N.}~\bibnamefont {Mizuochi}},\ and\
  \bibinfo {author} {\bibfnamefont {K.}~\bibnamefont {Semba}},\ }\bibfield
  {title} {\bibinfo {title} {{Coherent coupling of a superconducting flux qubit
  to an electron spin ensemble in diamond}},\ }\href
  {https://doi.org/10.1038/nature10462} {\bibfield  {journal} {\bibinfo
  {journal} {Nature}\ }\textbf {\bibinfo {volume} {478}},\ \bibinfo {pages}
  {221} (\bibinfo {year} {2011})}\BibitemShut {NoStop}%
\bibitem [{\citenamefont {Samkharadze}\ \emph {et~al.}(2016)\citenamefont
  {Samkharadze}, \citenamefont {Bruno}, \citenamefont {Scarlino}, \citenamefont
  {Zheng}, \citenamefont {DiVincenzo}, \citenamefont {DiCarlo},\ and\
  \citenamefont {Vandersypen}}]{Samkharadze2016}%
  \BibitemOpen
  \bibfield  {author} {\bibinfo {author} {\bibfnamefont {N.}~\bibnamefont
  {Samkharadze}}, \bibinfo {author} {\bibfnamefont {A.}~\bibnamefont {Bruno}},
  \bibinfo {author} {\bibfnamefont {P.}~\bibnamefont {Scarlino}}, \bibinfo
  {author} {\bibfnamefont {G.}~\bibnamefont {Zheng}}, \bibinfo {author}
  {\bibfnamefont {D.~P.}\ \bibnamefont {DiVincenzo}}, \bibinfo {author}
  {\bibfnamefont {L.}~\bibnamefont {DiCarlo}},\ and\ \bibinfo {author}
  {\bibfnamefont {L.~M.~K.}\ \bibnamefont {Vandersypen}},\ }\bibfield  {title}
  {\bibinfo {title} {{High-Kinetic-Inductance Superconducting Nanowire
  Resonators for Circuit QED in a Magnetic Field}},\ }\href
  {https://doi.org/10.1103/PhysRevApplied.5.044004} {\bibfield  {journal}
  {\bibinfo  {journal} {Physical Review Applied}\ }\textbf {\bibinfo {volume}
  {5}},\ \bibinfo {pages} {044004} (\bibinfo {year} {2016})}\BibitemShut
  {NoStop}%
\bibitem [{\citenamefont {Niepce}\ \emph {et~al.}(2019)\citenamefont {Niepce},
  \citenamefont {Burnett},\ and\ \citenamefont {Bylander}}]{Niepce2019a}%
  \BibitemOpen
  \bibfield  {author} {\bibinfo {author} {\bibfnamefont {D.}~\bibnamefont
  {Niepce}}, \bibinfo {author} {\bibfnamefont {J.}~\bibnamefont {Burnett}},\
  and\ \bibinfo {author} {\bibfnamefont {J.}~\bibnamefont {Bylander}},\
  }\bibfield  {title} {\bibinfo {title} {{High Kinetic Inductance NbN Nanowire
  Superinductors}},\ }\href {https://doi.org/10.1103/PhysRevApplied.11.044014}
  {\bibfield  {journal} {\bibinfo  {journal} {Physical Review Applied}\
  }\textbf {\bibinfo {volume} {10}},\ \bibinfo {pages} {044014} (\bibinfo
  {year} {2019})}\BibitemShut {NoStop}%
\bibitem [{\citenamefont {Maze}\ \emph {et~al.}(2011)\citenamefont {Maze},
  \citenamefont {Gali}, \citenamefont {Togan}, \citenamefont {Chu},
  \citenamefont {Trifonov}, \citenamefont {Kaxiras},\ and\ \citenamefont
  {Lukin}}]{Maze2011}%
  \BibitemOpen
  \bibfield  {author} {\bibinfo {author} {\bibfnamefont {J.~R.}\ \bibnamefont
  {Maze}}, \bibinfo {author} {\bibfnamefont {A.}~\bibnamefont {Gali}}, \bibinfo
  {author} {\bibfnamefont {E.}~\bibnamefont {Togan}}, \bibinfo {author}
  {\bibfnamefont {Y.}~\bibnamefont {Chu}}, \bibinfo {author} {\bibfnamefont
  {A.}~\bibnamefont {Trifonov}}, \bibinfo {author} {\bibfnamefont
  {E.}~\bibnamefont {Kaxiras}},\ and\ \bibinfo {author} {\bibfnamefont {M.~D.}\
  \bibnamefont {Lukin}},\ }\bibfield  {title} {\bibinfo {title} {{Properties of
  nitrogen-vacancy centers in diamond: the group theoretic approach}},\ }\href
  {https://doi.org/10.1088/1367-2630/13/2/025025} {\bibfield  {journal}
  {\bibinfo  {journal} {New Journal of Physics}\ }\textbf {\bibinfo {volume}
  {13}},\ \bibinfo {pages} {025025} (\bibinfo {year} {2011})}\BibitemShut
  {NoStop}%
\bibitem [{\citenamefont {Barson}\ \emph {et~al.}(2019)\citenamefont {Barson},
  \citenamefont {Krausz}, \citenamefont {Manson},\ and\ \citenamefont
  {Doherty}}]{Barson2019}%
  \BibitemOpen
  \bibfield  {author} {\bibinfo {author} {\bibfnamefont {M.~S.~J.}\
  \bibnamefont {Barson}}, \bibinfo {author} {\bibfnamefont {E.}~\bibnamefont
  {Krausz}}, \bibinfo {author} {\bibfnamefont {N.~B.}\ \bibnamefont {Manson}},\
  and\ \bibinfo {author} {\bibfnamefont {M.~W.}\ \bibnamefont {Doherty}},\
  }\bibfield  {title} {\bibinfo {title} {{The fine structure of the neutral
  nitrogen-vacancy center in diamond}},\ }\href
  {https://doi.org/10.1515/nanoph-2019-0142} {\bibfield  {journal} {\bibinfo
  {journal} {Nanophotonics}\ }\textbf {\bibinfo {volume} {8}},\ \bibinfo
  {pages} {1985} (\bibinfo {year} {2019})}\BibitemShut {NoStop}%
\bibitem [{\citenamefont {Jahnke}\ \emph {et~al.}(2015)\citenamefont {Jahnke},
  \citenamefont {Sipahigil}, \citenamefont {Binder}, \citenamefont {Doherty},
  \citenamefont {Metsch}, \citenamefont {Rogers}, \citenamefont {Manson},
  \citenamefont {Lukin},\ and\ \citenamefont {Jelezko}}]{Jahnke2015}%
  \BibitemOpen
  \bibfield  {author} {\bibinfo {author} {\bibfnamefont {K.~D.}\ \bibnamefont
  {Jahnke}}, \bibinfo {author} {\bibfnamefont {A.}~\bibnamefont {Sipahigil}},
  \bibinfo {author} {\bibfnamefont {J.~M.}\ \bibnamefont {Binder}}, \bibinfo
  {author} {\bibfnamefont {M.~W.}\ \bibnamefont {Doherty}}, \bibinfo {author}
  {\bibfnamefont {M.}~\bibnamefont {Metsch}}, \bibinfo {author} {\bibfnamefont
  {L.~J.}\ \bibnamefont {Rogers}}, \bibinfo {author} {\bibfnamefont {N.~B.}\
  \bibnamefont {Manson}}, \bibinfo {author} {\bibfnamefont {M.~D.}\
  \bibnamefont {Lukin}},\ and\ \bibinfo {author} {\bibfnamefont
  {F.}~\bibnamefont {Jelezko}},\ }\bibfield  {title} {\bibinfo {title}
  {{Electron-phonon processes of the silicon-vacancy centre in diamond}},\
  }\href {https://doi.org/10.1088/1367-2630/17/4/043011} {\bibfield  {journal}
  {\bibinfo  {journal} {New Journal of Physics}\ }\textbf {\bibinfo {volume}
  {17}},\ \bibinfo {pages} {043011} (\bibinfo {year} {2015})}\BibitemShut
  {NoStop}%
\bibitem [{\citenamefont {Krantz}\ \emph {et~al.}(2019)\citenamefont {Krantz},
  \citenamefont {Kjaergaard}, \citenamefont {Yan}, \citenamefont {Orlando},
  \citenamefont {Gustavsson},\ and\ \citenamefont {Oliver}}]{Krantz2021}%
  \BibitemOpen
  \bibfield  {author} {\bibinfo {author} {\bibfnamefont {P.}~\bibnamefont
  {Krantz}}, \bibinfo {author} {\bibfnamefont {M.}~\bibnamefont {Kjaergaard}},
  \bibinfo {author} {\bibfnamefont {F.}~\bibnamefont {Yan}}, \bibinfo {author}
  {\bibfnamefont {T.~P.}\ \bibnamefont {Orlando}}, \bibinfo {author}
  {\bibfnamefont {S.}~\bibnamefont {Gustavsson}},\ and\ \bibinfo {author}
  {\bibfnamefont {W.~D.}\ \bibnamefont {Oliver}},\ }\bibfield  {title}
  {\bibinfo {title} {{A quantum engineer's guide to superconducting qubits}},\
  }\href {https://doi.org/10.1063/1.5089550} {\bibfield  {journal} {\bibinfo
  {journal} {Applied Physics Reviews}\ }\textbf {\bibinfo {volume} {6}},\
  \bibinfo {pages} {021318} (\bibinfo {year} {2019})}\BibitemShut {NoStop}%
\bibitem [{\citenamefont {Kim}\ \emph {et~al.}(2023)\citenamefont {Kim},
  \citenamefont {Kurokawa}, \citenamefont {Sakai}, \citenamefont {Koshino},
  \citenamefont {Kosaka},\ and\ \citenamefont {Nomura}}]{Kim2023b}%
  \BibitemOpen
  \bibfield  {author} {\bibinfo {author} {\bibfnamefont {B.}~\bibnamefont
  {Kim}}, \bibinfo {author} {\bibfnamefont {H.}~\bibnamefont {Kurokawa}},
  \bibinfo {author} {\bibfnamefont {K.}~\bibnamefont {Sakai}}, \bibinfo
  {author} {\bibfnamefont {K.}~\bibnamefont {Koshino}}, \bibinfo {author}
  {\bibfnamefont {H.}~\bibnamefont {Kosaka}},\ and\ \bibinfo {author}
  {\bibfnamefont {M.}~\bibnamefont {Nomura}},\ }\bibfield  {title} {\bibinfo
  {title} {{Diamond optomechanical cavity with a color center for coherent
  microwave-to-optical quantum interfaces}},\ }\href
  {https://doi.org/10.1103/PhysRevApplied.20.044037} {\bibfield  {journal}
  {\bibinfo  {journal} {Physical Review Applied}\ }\textbf {\bibinfo {volume}
  {20}},\ \bibinfo {pages} {044037} (\bibinfo {year} {2023})}\BibitemShut
  {NoStop}%
\end{thebibliography}
\end{document}